\documentclass[%
reprint,
 notitlepage,11pt,onecolumn,tightenlines,
 longbibliography,
 amsmath,amssymb,
 aps,
 pre,
 nofootinbib
]{revtex4-1}

\usepackage{graphicx}
\usepackage{dcolumn}
\usepackage{bm}
\usepackage{xcolor} 
\usepackage{enumitem} 

\begin{document}

\title{Minimal descriptions of cyclic memories}

\author{Joseph D. Paulsen$^{1,}$}
 \email{jdpaulse@syr.edu}
 \affiliation{Department of Physics and Soft and Living Matter Program, Syracuse University, Syracuse, NY 13244, USA}
\affiliation{Kavli Institute for Theoretical Physics, Santa Barbara, CA 93106, USA}
\author{Nathan C.\ Keim$^{1,}$}
 \email{nkeim@calpoly.edu \\ $^1$These authors contributed equally to this work.}
 \affiliation{Department of Physics, California Polytechnic State University, San Luis Obispo, CA 93407}
\affiliation{Kavli Institute for Theoretical Physics, Santa Barbara, CA 93106, USA}

\date{\today}

\begin{abstract}
Many materials that are out of equilibrium can ``learn'' one or more inputs that are repeatedly applied. 
Yet, a common framework for understanding such memories is lacking. 
Here we construct minimal representations of cyclic memory behaviors as directed graphs, and we construct simple physically-motivated models that produce the same graph structures. 
We show how a model of worn grass between park benches can produce multiple transient memories---a behavior previously observed in dilute suspensions of particles and charge-density-wave conductors---and the Mullins effect. 
Isolating these behaviors in our simple model allows us to assess the necessary ingredients for these kinds of memory, and to quantify memory capacity. 
We contrast these behaviors with a simple Preisach model that produces return-point memory. 
Our analysis provides a unified method for comparing and diagnosing cyclic memory behaviors across different materials. 
\end{abstract}

\maketitle

\section{Introduction}\label{introduction}
Materials that are out of equilibrium can sometimes form memories of their past. 
Rubber and rocks may remember the largest loading applied to them~\cite{Mullins48,Kurita79,Schmoller13}; glasses may remember aspects of their relaxation~\cite{Jonason98,Zou10,Gilbert15,Yang17}; a sheet of plastic can remember how severely~\cite{Matan02} or how long~\cite{Lahini17} it was crumpled. 
In each of these systems, information may be stored and then retrieved at a later time if there is some established protocol for doing so. 
Despite the simplicity of this idea and the many common features shared by diverse examples~\cite{Keim18review}, there is presently no overarching framework for understanding memories in materials. 

One promising place to start building such a framework is in systems where the driving may be divided into cycles. 
Examples include a repeatedly sheared granular material or amorphous solid~\cite{Toiya04,Royer15,Fiocco14,Keim14}, or a set of magnetic domains in an oscillating external field~\cite{Barker83,Sethna93}. 
Here we distill the essential aspects of several cyclic memory behaviors into simple transition graphs, which represent the different memory-encoding macrostates and the transitions between them.
We show that this is a succinct and powerful way to compare these various behaviors, and we highlight how this approach can help diagnose memory behaviors in experiments.

Some physical systems lend themselves naturally to a graph representation~\cite{Mungan18} because they are clearly discrete (e.g., spin systems), while others are just beginning to be described in this way (e.g., amorphous solids under quasistatic shear~\cite{Fiocco15,Mungan18}).
One question that arises in this effort is whether a behavior called \emph{multiple transient memory} (MTM) may be captured with a small set of discrete states. 
This behavior is observed in charge-density wave conductors' memory of electrical pulse duration \cite{Coppersmith97, Povinelli99} and non-Brownian suspensions' memory of the amplitude of oscillatory shear \cite{Keim11, Keim13, Paulsen14, Adhikari18}. 
When these systems are driven cyclically, 
they self-organize into steady states that store the repeated value (i.e., the pulse duration or the strain amplitude $\gamma_0$). 
Moreover, when driven with multiple amplitudes on successive cycles, they display the following properties: (1) during the transient, multiple $\gamma_i$ may be encoded; (2) the order in which the values are applied is not crucial---introducing a new \(\gamma_i\) during the transient may degrade the memories of previous values but does not erase them; (3) when a steady state is reached, it can only retain memories of the smallest and largest \(\gamma_i\) that were applied; and remarkably, (4) a small amount of noise allows all memories to be retained indefinitely \cite{Coppersmith97, Povinelli99, Keim11, Keim13, Paulsen14}.

\begin{figure}[t]
\centering
\includegraphics[scale=0.55]{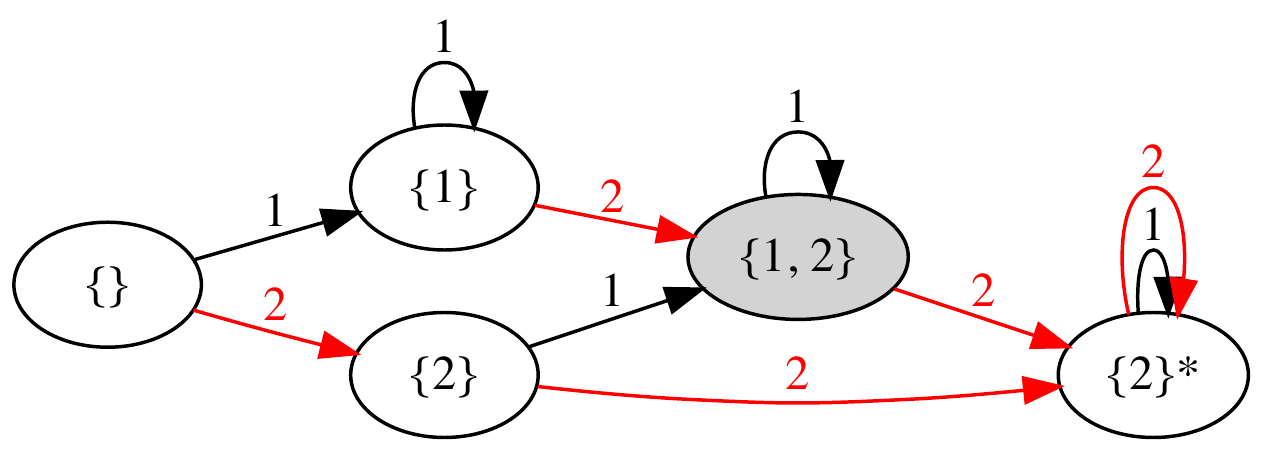}
\caption{\label{fig:minimal} 
Minimal transition graph exhibiting multiple transient memory, starting from a featureless state. 
States are labeled with the memories they store; arrows are labeled with the amplitude applied during a given cycle ($\gamma=1$ or $2$) and point to the resulting state. 
The state $\{2\}^*$ cannot have a memory of $1$ written in it.
}
\end{figure}

Our approach is to consider small sets of discrete states that obey a given memory behavior. 
Figure~\ref{fig:minimal} shows five states and transitions that exhibit properties 1--3 of MTM, where the states are labeled with the memories they store. 
The system starts in a memoryless state, $\{\}$. 
During each cycle an amplitude of either $\gamma_0=1$ or $2$ is applied. 
Hence, there are two arrows emanating from each state, labeled with the driving amplitude for that transition. 
(An arrow may point to the state where it started if the driving does not change the state.) 

Consider first the driving sequence: $\gamma_0 = 2,1,2,1,2,1,\ldots$. 
Following the transitions shown in Fig.~\ref{fig:minimal}, this leads to a series of memory states: $\{2\},\{1, 2\},\{2\}^*,\{2\}^*,\{2\}^*,\{2\}^*,\ldots$, where the absorbing state is denoted with a $^*$ to indicate that it is incapable of having a memory of $1$ written in it. 
The state $\{1, 2\}$ is obtained once, demonstrating that multiple memories may be encoded in the transient (Property 1). 
To see that Property 2 holds, one may consider a different driving sequence: $\gamma_0 = 1,2,1,2,1,2,\ldots$, which also reaches the state $\{1, 2\}$ during the transient. 
Any repeated sequence containing $\gamma_0=1$ and $2$ eventually leads to the absorbing state $\{2\}^*$, satisfying Property 3. 
(For simplicity, we do not denote the memory of the smallest input, which for a suspension driven cyclically between $\gamma=0$ and $\gamma_i$ corresponds to a ``trivial'' memory written at $\gamma=0$~\cite{Paulsen14}; here this memory would be present in all but the featureless state, $\{\}$.)

While this description is useful in demonstrating properties 1--3 of MTM in a minimal set of discrete states, it is somewhat artificial; we did not provide a physical reason for this arrangement of states or the transitions. 
Thus, in this paper we also describe a novel, simple, and physically-motivated model proposed by Sidney Nagel called the ``park bench model,'' which captures the distinctive aspects of multiple transient memory. 
We then show how noise may be introduced into the model to prolong the transient period indefinitely.  Our analysis of noise in this model allows us to define a memory capacity---the number of distinct memories that can be retained simultaneously---which has been elusive in other systems with MTM. 
We also demonstrate that this model's behavior may reduce to a simpler type of memory (the Mullins effect) for some initial states.
To show the versatility of our approach and highlight differences from MTM, we then construct minimal graphs of return-point memory (RPM)~\cite{Sethna93,Lilly93,Deutsch04,Ortin11}, and we describe a simple physical model that produces this memory structure. 
Finally, we describe how this graph framework can help suggest specific hypotheses and tests for experiments and simulations, which should be useful in systems where the distinctions among the memory behaviors are not as clear~\cite{Fiocco14,Fiocco15,Bachelard17,Dobroka17,Adhikari18}.
These results are a concrete step towards developing a broad organizing framework for memories in matter.

\section{Results}
\subsection{Park bench model}\label{sec:bench}
Consider a lawn with several benches arranged in a straight line. 
As visitors walk from the end of the park to any one of the benches, they gradually wear a path into the grass. 
As an observer, what can you deduce about previous visitors by looking at the grass? 
If the worn path ends at one of the benches, you may infer that many people stopped at that particular bench. 
On a finer scale, if the grass is somewhat worn leading up to the second bench but \textit{more} worn up to the first bench, you might infer that some visitors walked to only the first bench and others continued on to the second bench. 
Thus, any spatial variation in the wear provides information about the past. 

Perhaps counter-intuitively, information may also be \textit{lost} through wear. 
Suppose the interval from the entrance to the second bench is so thoroughly traveled that the grass is completely worn down to the soil. 
In that case, you lose any clue that the first bench was visited at all; there can be no change in the state of the grass along an interval if it is worn to its roots. 
Even more behaviors are possible if the grass is gradually growing back at all times; we consider this possibility in Section~\ref{sec:noise}.

\begin{figure}[b]
\centering
\raisebox{1.17in}{\textbf{a)}} \raisebox{0.9\height}{\includegraphics[scale=0.5]{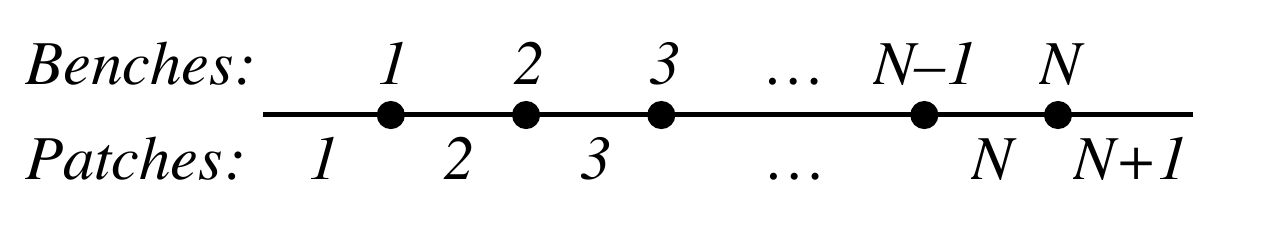}} 
\raisebox{1.17in}{\textbf{b)}} \includegraphics[scale=0.55]{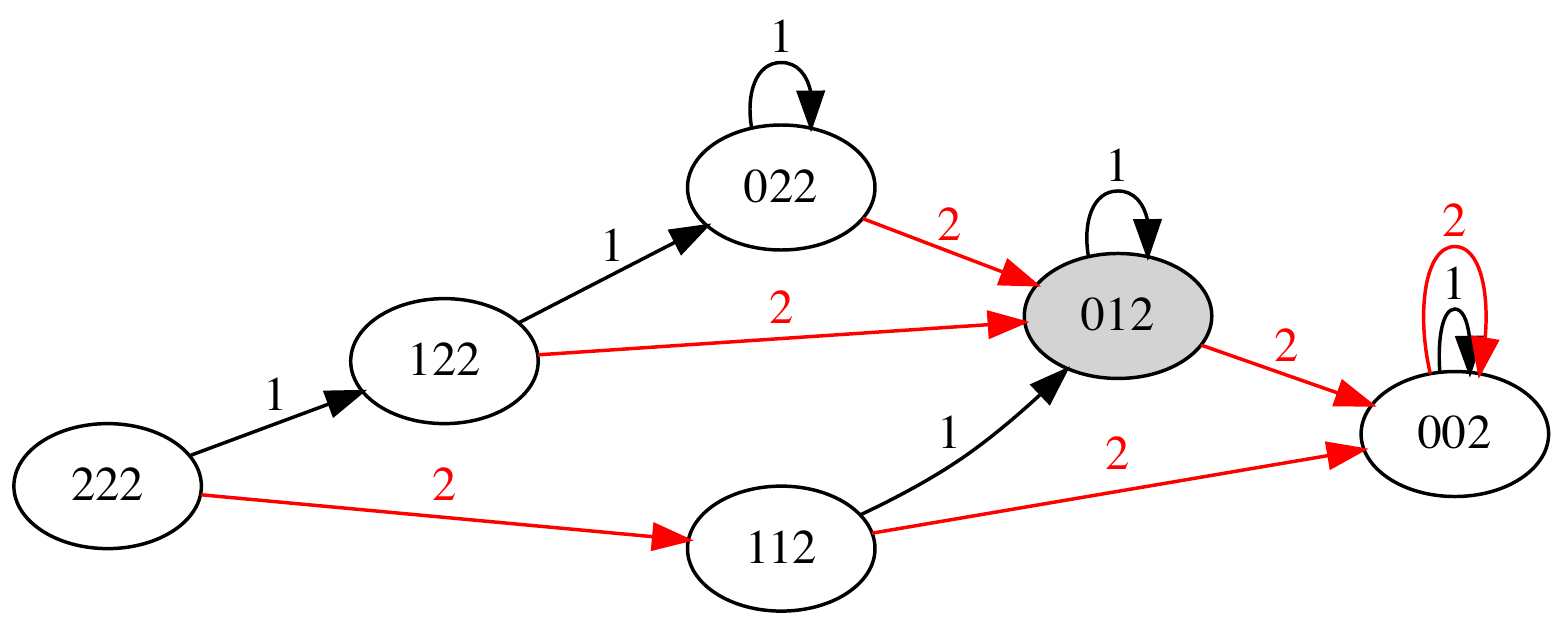}
\caption{\label{fig:bench23} 
Park bench model. 
\textbf{(a)} Arrangement of benches and patches. 
\textbf{(b)} Transition graph with $N=2$, $h_\text{init}=2$. The digits in each node indicate the grass height on each patch. 
Each edge is labeled with a number $i$, representing an excursion from the leftmost position to the $i$th bench and back. 
Shaded states encode multiple memories.}
\end{figure}

To make these notions precise, we consider a one-dimensional model with $N$ benches separating $N+1$ patches of grass on a line, as drawn in Fig.~\ref{fig:bench23}a. 
Each patch of grass has initial height $h_\text{init}$. 
During a cycle, a visitor starts at the park entrance, walks to the $n$th bench (thus passing all the benches before it) and then returns to the park entrance. 
As a result, the grass height decreases by one unit on patches $1$ through $n$. 
We consider cyclic driving where patrons visit any sequence of benches in this manner. 
We denote the state of the system by a string of $N+1$ integers that record the grass height on each patch, including the last, inaccessible patch. 
(The benches may be visualized as sitting between the integers in the string.) 
A valid state is thus given by a nondecreasing string of length $N+1$, of any values $0$ through $h_\text{init}$, ending with $h_\text{init}$ for the last patch.

\begin{figure*}
\centering
\includegraphics[scale=0.55]{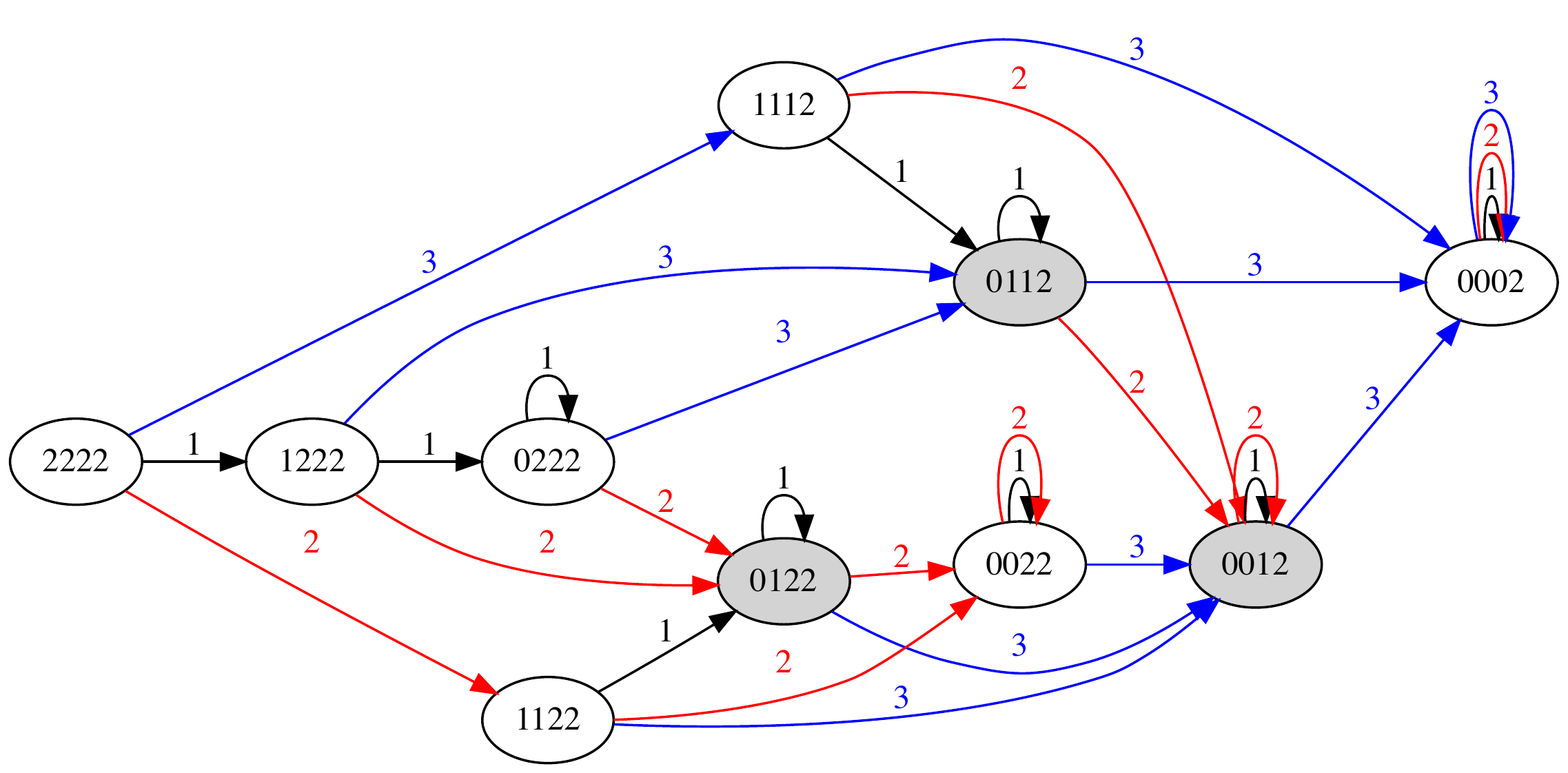}
\caption{\label{fig:bench33} Transition graph for the park bench model with $N=3$, $h_\text{init}=2$.}
\end{figure*}

\textit{Graph representation and behavior ---}
To show all the possible ways a system may evolve, we enumerate the accessible states and draw a directed graph of the transitions between them, as shown in Fig.~\ref{fig:bench23}b for the system $N=2$, $h_\text{init}=2$. 
Each state has $N$ arrows coming out of it (here $N=2$), representing the $N$ possible amplitudes of a park-goer's stroll to any of the $N$ benches and back. 
The state with $N$ zeros and a single pristine patch of height $h_\text{init}$ represents a completely worn path up to the last bench. 
This is a fixed point of the driving; all arrows from this state point back to it. 

Except for the initial state $222$, all the states store some amount of memory. 
States $002$ and $112$ are uniformly worn up to the second bench, so they store a memory of only the second bench. 
(Although it is possible that trips to the first bench were also involved in reaching state $002$, there is no way to know that from these grass heights.) 
The state $012$ stores two memories: it implies that both the first and second benches were visited. 
By considering these states in terms of their memory content, one may readily verify that this graph has the same structure as the minimal graph for MTM shown in Fig.~\ref{fig:minimal} (plus the additional state $022$). 
Likewise, by considering the driving sequences $\gamma = 2,1,2,1,\ldots$ and $\gamma = 1,2,1,2,\ldots$, one can check properties 1--3 of MTM\footnote{For simplicity, we neglect the memory of the smallest input, which here corresponds to a ``trivial'' memory written at $\gamma=0$. One could add benches and patches to the left of patch 1 and denote also the memories of the smallest inputs in the transition graphs, but the basic memory behavior would be the same.}. 

Figure~\ref{fig:bench33} shows the transition graph for the system with $N=3$, $h_\text{init}=2$. 
Here, three states contain multiple memories: $0122$ stores memories of $1$ and $2$; $0112$ stores memories of $1$ and $3$; and $0012$ stores memories of $2$ and $3$. 
The values of $N$ and $h_\text{init}$ set the memory capacity of the system: in the example of Fig.~\ref{fig:bench33}, the initial grass height is not tall enough to store memories of $1$, $2$, and $3$ simultaneously. 
In general, the model can store at most $\min(N,h_\text{init})$ memories at one time. 

\textit{Cyclic memory behaviors as properties of transition graphs ---} 
Properties 1--3 of MTM may be checked on an arbitrary transition graph to diagnose its memory behavior. 
Property 1 says there should be a state with multiple memories that may be reached from the initial state. 
Property 2 says this state should be reachable by applying the amplitudes in any order. 
Property 3 says that the fixed point of any repeated driving sequence should store just one memory. 
Importantly, these properties may be checked by examining the graph structure without any reference to the physics that produced the graph.

\subsection{Addition of noise}\label{sec:noise}

We now consider Property 4 of MTM in the park bench model. 
Charge-density wave conductors~\cite{Povinelli99} and non-Brownian suspensions~\cite{Keim11,Keim13} have the remarkable property that noise \emph{enhances} memory retention by preventing the system from reaching the final absorbing state. 
In the charge-density wave model, this is accomplished by resetting a few randomly-chosen elastic bonds on each cycle; in the suspension, by small random displacements of every particle. 
To perturb the grass, we assign each patch a small probability of increasing its height by 1 unit. 
Under sustained driving we want the system to reach a fluctuating equilibrium state, so shorter grass should grow faster. 
A simple and suitable form for the probability of a patch to grow in each cycle is:
\begin{equation}\label{pgrow}
p(h_i \to h_i + 1) = \alpha \frac{h_\text{init}-h_i}{h_\text{init}},
\end{equation}
where $h_i$ is the present height of the $i$th patch of grass 
and $\alpha$ controls the amount of noise. 
We apply the noise at the beginning of each cycle, before driving.
Because the grass may grow anywhere, adding noise to the model leads to newly-accessible states where grass heights do not necessarily increase from left to right. 
The proliferation of new states and transitions leads us to distinguish this memory behavior as MTM with noise (MTMN).

Focusing on a single patch of grass, we can predict its steady-state height, $\overline{h_i(t)}$, in the case where there is almost always some grass to remove ($h_i(t) > 0$ at long times)\footnote{In particular, this analysis requires $\alpha > \overline{D_i}$, otherwise the assumption $h_i(t) > 0$ is frequently violated at patch $i$.}. 
Such a steady state is reached when the average growth rate matches the time-averaged driving:
\begin{equation}\label{balance}
\alpha \frac{h_\text{init}- \overline{h_i(t)} }{h_\text{init}} = \overline{D_i} ,
\end{equation}
where 
$\overline{D_i}$ has value 1 if the $i$th patch is visited on each cycle, $\frac{1}{2}$ if it is visited on alternate cycles, etc. 
Thus, $\overline{h_i(t)} = ( 1 - \overline{D_i}/\alpha ) h_\text{init}$. 
This is a stable fixed point; a positive (or negative) fluctuation leads to a slight decrease (or increase) in the growth rate, because of the sign of $\overline{h_i(t)}$ in Eq.~\ref{balance}. 
The existence of an equilibrium height that depends on the local driving at each patch is what allows the system to store multiple memories in the steady state. Namely, when driving with multiple amplitudes, $\overline{D_i}$ will vary from patch to patch, and the $\overline{h_i}$ observed at each patch will encode that variation. 
Memories of the reversal points (i.e., the amplitudes of the driving cycles) are thus stored as the locations of jumps in the steady-state height $h_i(t)$ as a function of patch number, $i$.

\begin{figure}[t]
\centering
\raisebox{2.01in}{\textbf{a)}}
\includegraphics[scale=0.59]{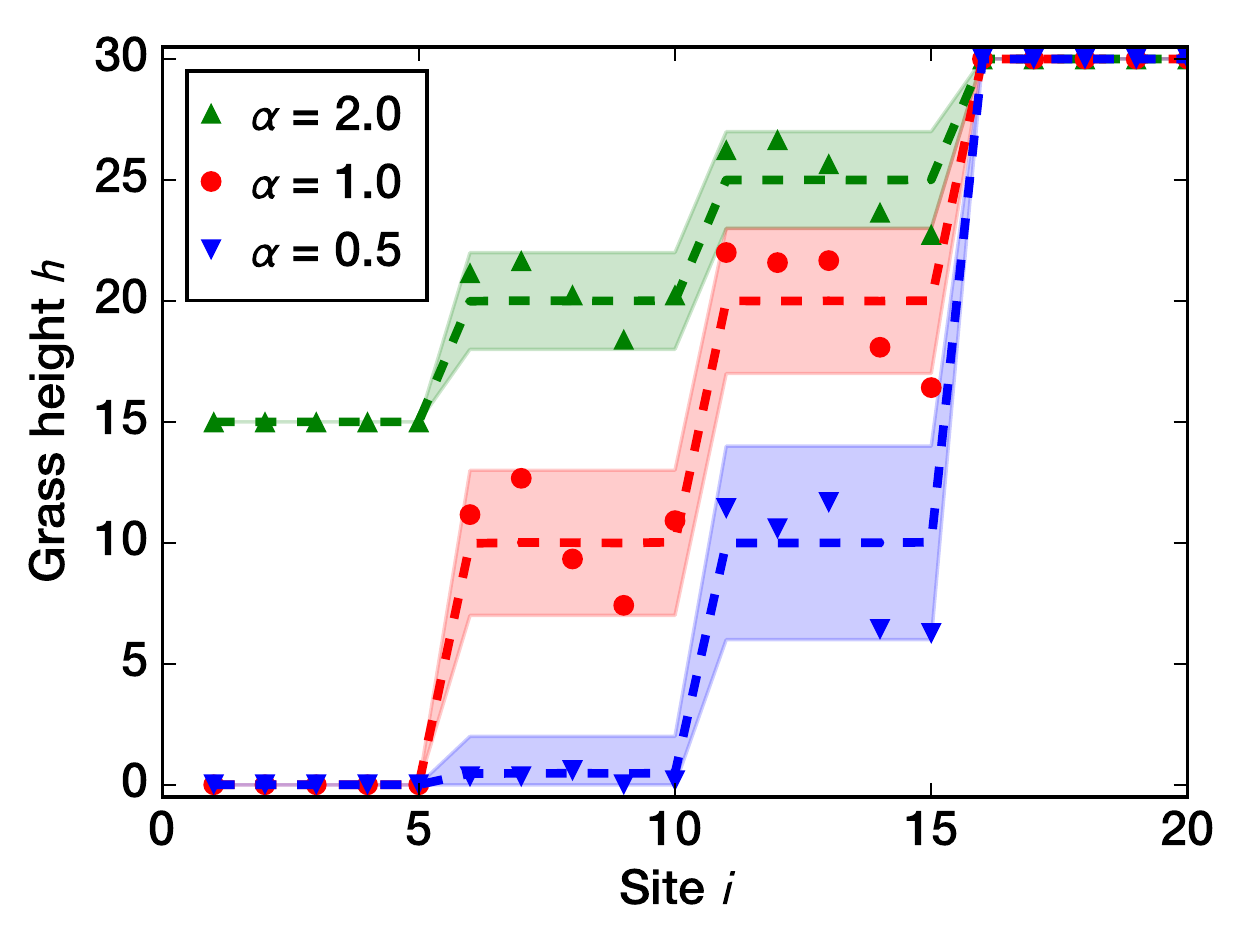} 
\hspace{0.1in}
\raisebox{2.01in}{\textbf{b)}} 
\includegraphics[scale=0.59]{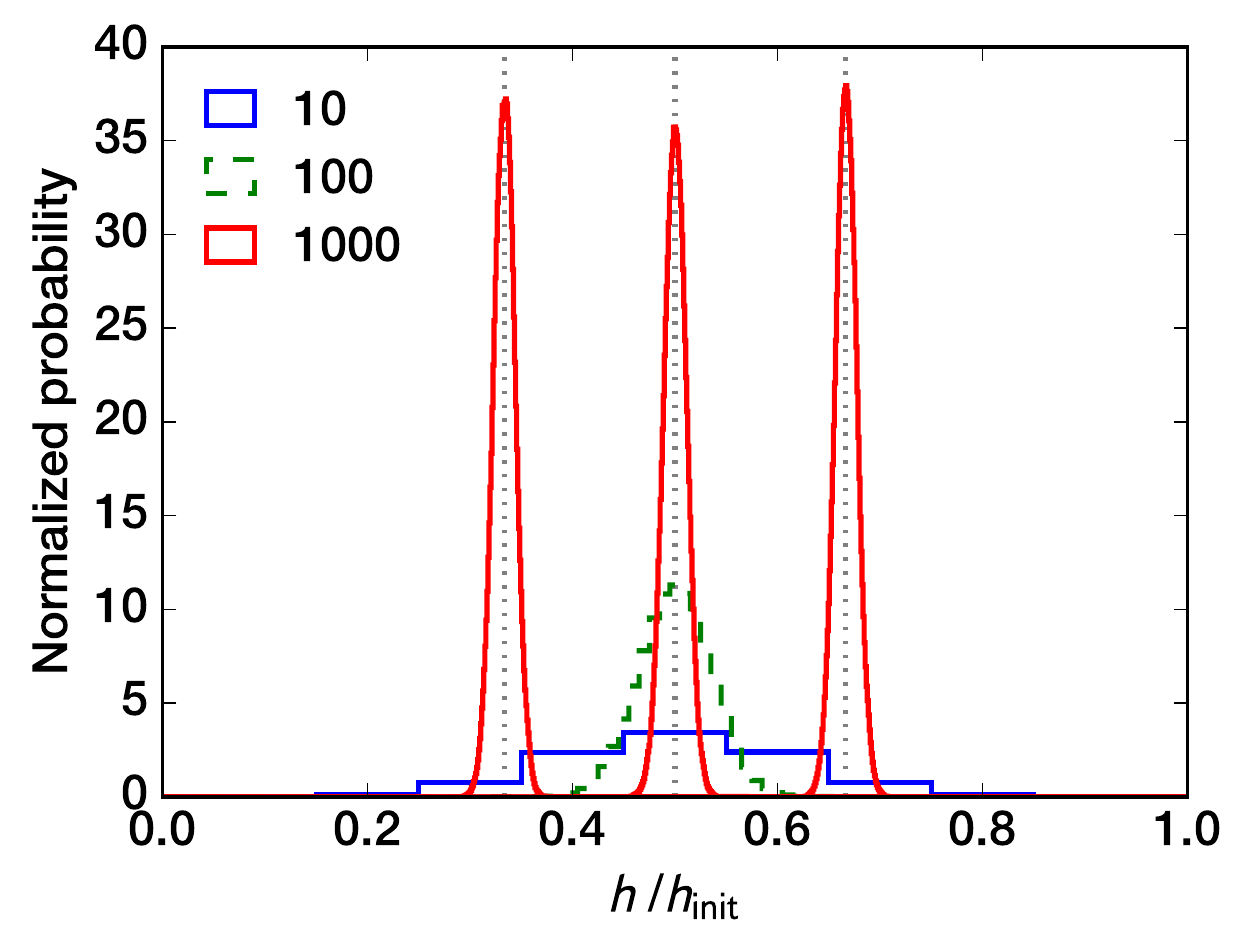}
\caption{\label{fig:noisy} 
Grass heights in the steady state with noise.
\textbf{(a)} Simulations with $N = 20$, $h_\text{init} = 30$, and $\alpha = 0.5, 1.0, 2.0$. The system is driven with the repeating pattern: $\gamma=15,10,5,15,10,5,\dots$. Points show averages over the last 12 cycles of the simulation. Dashed lines show averages over $10^6$ cycles. Shaded regions show the size of fluctuations, bounded by the 5th and 95th percentiles of values. In each case, all three memories are apparent in the steady state, though with $\alpha = 1.0$ they are most distinct and there is the least chance of a momentary lapse of memory.
\textbf{(b)} Steady-state distributions of $h/h_\text{init}$ of a single site simulated for $10^7$ cycles, with $\alpha = 1.0$ and $h_\text{init} = 10, 100, 1000$ (see legend), for $\overline{D}=1/2$. Distributions are also shown for $h_\text{init}=1000$ and $\overline{D}=1/3$ and $2/3$ (peaked near $h/h_\text{init} = 2/3$ and $1/3$, respectively, which are indicated with dotted vertical lines). Each curve is normalized to have area 1.
}
\end{figure}

To demonstrate these behaviors, we simulate a system with $N=20$ and $h_\text{init}=30$. 
Figure~\ref{fig:noisy}a shows the average steady-state grass heights over $10^6$ cycles and their fluctuations, for three memories at different values of $\alpha$. (We begin the simulation with a transient of $10^6$ cycles that is not recorded.) 
As in other systems with MTM~\cite{Povinelli99,Keim13}, the noise does not require fine-tuning; we find that for a wide range of $1 \lesssim \alpha \ll h_\text{init}$, all memories will be preserved on average. 
Notably, the memory at $\gamma=5$ is even retained when $\alpha = 0.5$, despite the maximum growth rate being smaller than the average driving rate $\overline{D_i}$ at each patch $i \leq 10$. 
Nevertheless, the patches in the interval $6 \leq i \leq 10$ can fluctuate up to $h_i=1$ or higher at times, since they have a finite probability of gaining height on cycles with $\gamma = 5$. 
This stochastic case leads to $\overline{h_i(t)} > 0$, which allows the memory to persist.

\textit{Height fluctuations and memory capacity ---}
Equation~\ref{balance} for the mean grass height provides a basic framework for storing multiple memories in the noisy park bench model. 
However, understanding fluctuations is also of crucial importance for retrieving multiple memories, since noise may mask a memory or create a spurious one. 
One approach to reduce such errors when reading out the memories is to perform many measurements and average them together. 
Nonetheless, in the absence of such averaging, one wants to know the inherent limitations on storing and retrieving multiple memories. 
Intuitively, the plateaus in Fig.~\ref{fig:noisy}a must be separated by vertical steps that are larger than the characteristic size of the fluctuations. 
This consideration puts a sharp limit on the memory capacity of the system, since it tells how many discernible jumps in grass height may occur in a system with maximum height $h_\text{init}$.

To investigate these behaviors, we consider the behavior of a single patch of grass in the steady state. 
We measure the probability distribution of its height, $P(h_i)$, in simulations as shown in Fig.~\ref{fig:noisy}b.
Each simulation for $P(h_i)$ runs for $1.1 \times 10^7$ cycles, and we discard the initial $10^6$ cycles as a transient. 
The distribution is peaked about $\overline{h_i(t)}$ and approaches a smooth Gaussian as $h_\text{init}$ increases. 
These distributions indicate how well the steady-state height between two neighboring patches may be distinguished with a single observation of their instantaneous state; clearly this becomes easier with increasing $h_\text{init}$. 
Notably, the width of the distribution (characterized by its standard deviation, $\sigma$) does not change significantly when we vary the driving to $\overline{D_i}=1/3$ or $2/3$. 
Moreover, changing the driving pattern while keeping $\overline{D_i}$ fixed does not have a large effect on $\sigma$: The training patterns $\gamma=2,1,2,1,\ldots$ and $\gamma=2,2,1,1,2,2,1,1,\ldots$ both give $\sigma=11.16$ for $h_\text{init}=1000$, $\alpha=1$, whereas training with a 100-cycle pattern of 50 $2$'s followed by 50 $1$'s gives a slightly larger $\sigma=13.30$.

\begin{figure}[t]
\centering
\raisebox{2.03in}{\textbf{a)}}
\includegraphics[scale=0.65]{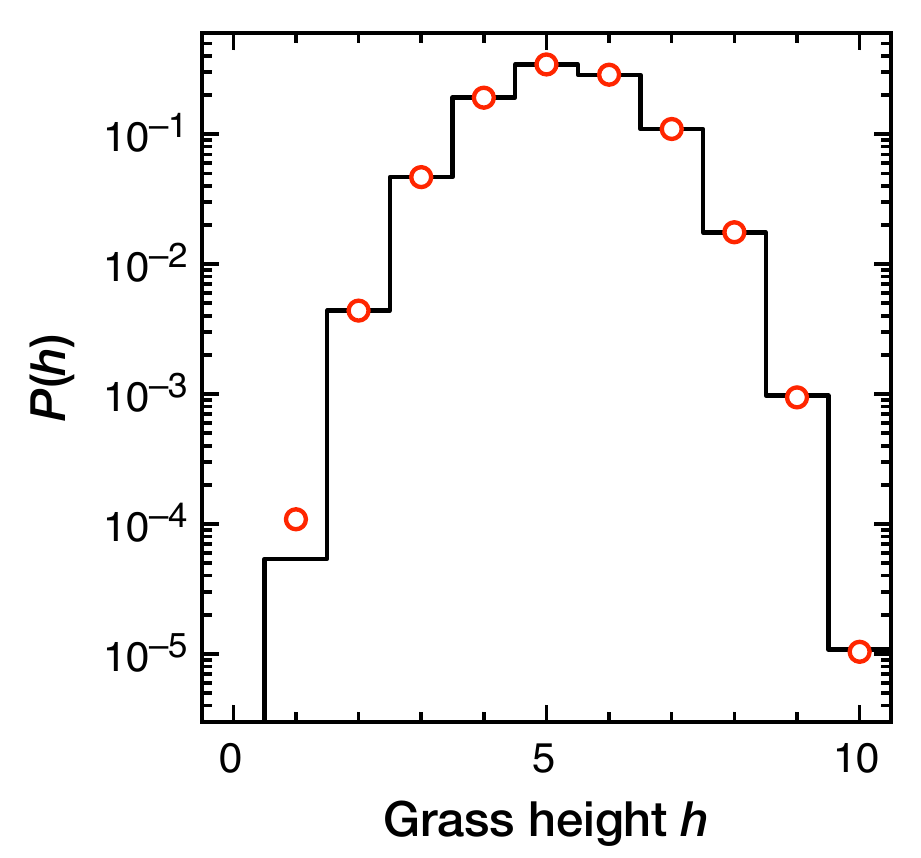}
\hspace{0.25in}
\raisebox{2.03in}{\textbf{b)}} \raisebox{-0.025in}{\includegraphics[scale=0.65]{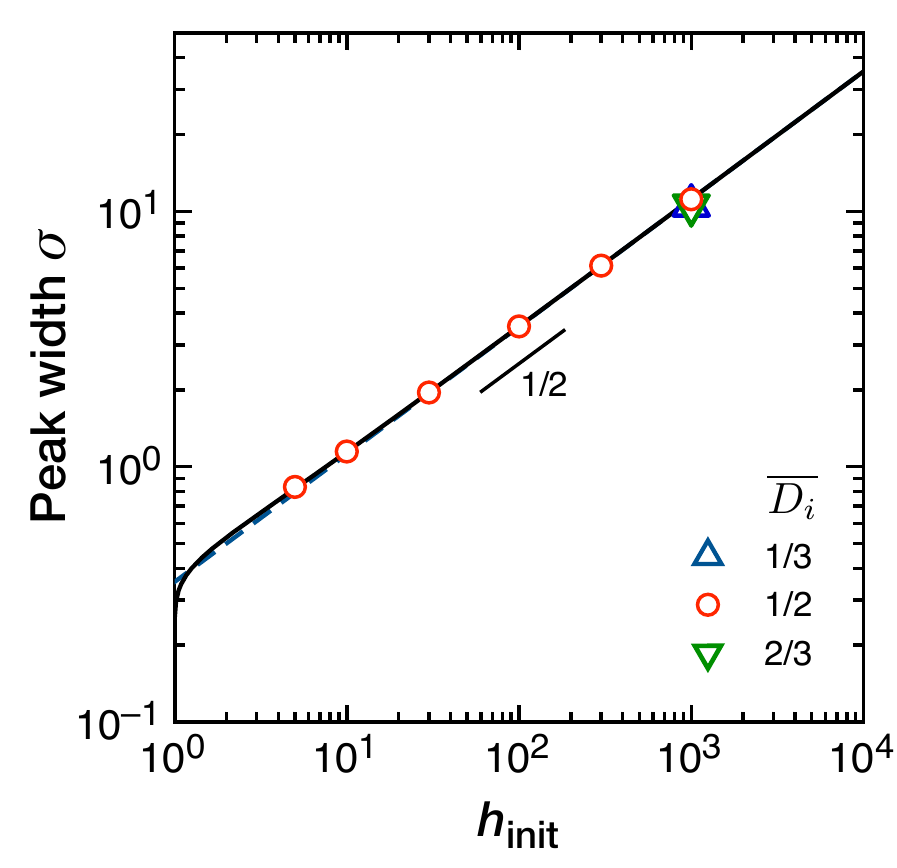}}
\caption{\label{fig:markov} 
Markov chain analysis of the park bench model with noise. 
\textbf{(a)} Steady-state probability distribution of grass height for $h_\text{init}=10$, $\alpha=1$, obtained by computing the transition matrix and finding the appropriate eigenvector (solid line). 
The values are in excellent agreement with the results from simulations (symbols). 
\textbf{(b)} Memory peak width, $\sigma$, versus $h_\text{init}$ for $\alpha=1$. 
Solid line: Result from Markov chain analysis, Eq.~\ref{sigma-scaling}.  
Dashed line: Asymptotic form, $\sqrt{h_\text{init}/8\alpha}$, which differs only slightly from the solid line at small $h_\text{init}$. 
Both curves capture the simulation results extremely well, for three values of the driving (symbols).
}
\end{figure}

\textit{Markov chain analysis ---}
We can quantitatively capture the above behaviors using discrete-time Markov chains. 
We focus on the evolution of $h_i(t)$ over a repeated training pattern, which may consist of multiple cycles of driving. 
We denote the probability of transitioning from height $a$ to height $b$ by a transition matrix $\mathbf{P}_{ab}$. 
Each element of this matrix may be constructed by applying Eq.~\ref{pgrow} to a unit probability starting in state $a$ and following its evolution over the entire training pattern. 
We consider the steady-state probability distribution $P(h)$ reached at long times, which we denote by a row vector $\pi$, with $\pi_a$ the probability of the grass having height $a$. 
This distribution is intimately related to the transition matrix: $\pi$ is an eigenvector of $\mathbf{P}_{ab}$ with eigenvalue $1$. 
For $h_\text{init}=10$, $\alpha=1$, we find exactly one such eigenvector. 
We show this distribution in Fig.~\ref{fig:markov}a, which is in excellent agreement with the steady-state probabilities observed in simulations.

To gain insight into how the width of the distribution, $\sigma$, depends on $h_\text{init}$ and $\alpha$, we consider the behavior near the most probable state, $a=a^*$. 
Balancing the probability flow out of and into $a^*$, we have:
\begin{equation}\label{inout}
\sum_{b \neq a^*} \pi_{a^*} \mathbf{P}_{a^*b} = \sum_{b \neq a^*} \pi_{b} \mathbf{P}_{ba^*}
\end{equation}
We then model the entries of $\pi$ as a Gaussian, so that $\pi_a = C \exp(-\frac{1}{2}(a^*-a)^2/\sigma^2)$. 
This reduces the number of degrees of freedom in modeling the 
vector $\pi$ from $\mathcal{O}(h_\text{init})$ to $1$ (i.e., the value of $\sigma$). 
Plugging this form into Eq.~\ref{inout} and supplying the values for the $\mathbf{P}_{ab}$ yields an equation for $\sigma$. 
For the simplest case of a two-cycle training pattern with $\overline{D_i} = 1/2$, this equation may be solved exactly to give:
\begin{equation}\label{sigma-scaling}
\sigma = \left(-2 \log\left(\frac{h_\text{init}(h_\text{init}-\alpha)}{(h_\text{init}+\alpha)(h_\text{init}+2\alpha)} \right)\right)^{-1/2} \sim \sqrt{\frac{h_\text{init}}{8\alpha}} \ \ \ (h_\text{init} \gg \alpha).
\end{equation}
Figure~\ref{fig:markov}b compares this prediction with measurements of $\sigma$ from simulations. 
The prediction with no fitting parameters captures the data for $\overline{D_i} = 1/2$ extremely well. 
Moreover, Eq.~\ref{sigma-scaling} gives a good estimate for the peak width for other training patterns, such as the three-cycle patterns with $\overline{D_i} = 1/3$ and $2/3$, where the Markov chain analysis involves significantly more terms. 
Thus, we obtain a rudimentary estimate of the memory capacity of $\min(N,\sqrt{h_\text{init}})$ for $\alpha=1$ (as compared with $\min(N,h_\text{init})$ for the case without noise, although in that case, multiple memories are impossible in the steady state). 
We note that a different scaling for $\sigma$ should arise when $\overline{D_i}$ is close to $0$ or $1$, which would add corrections to this estimate.
These results lay the groundwork for understanding the memory capacity of this system in a concrete way. Because the fluctuations are so nearly Gaussian, there is a viable basis for predicting the error rates of readout protocols involving multiple sites or averaging over time. This kind of precise understanding of memory capacity has been elusive in other systems that can store multiple memories under cyclic driving~\cite{Coppersmith97,Povinelli99,Keim11,Keim13,Paulsen14,Fiocco15,Adhikari18}.

\subsection{Recovering the Mullins effect}\label{sec:recover-Mullins}
\begin{figure}[h]
\centering
\raisebox{0.7in}{\textbf{a)}} 
\includegraphics[scale=0.55]{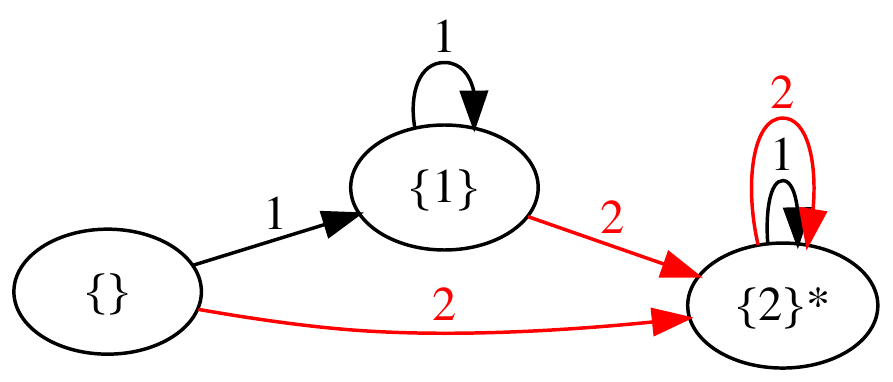} 
\hspace{0.5in}
\raisebox{0.7in}{\textbf{b)}} 
\includegraphics[scale=0.55]
{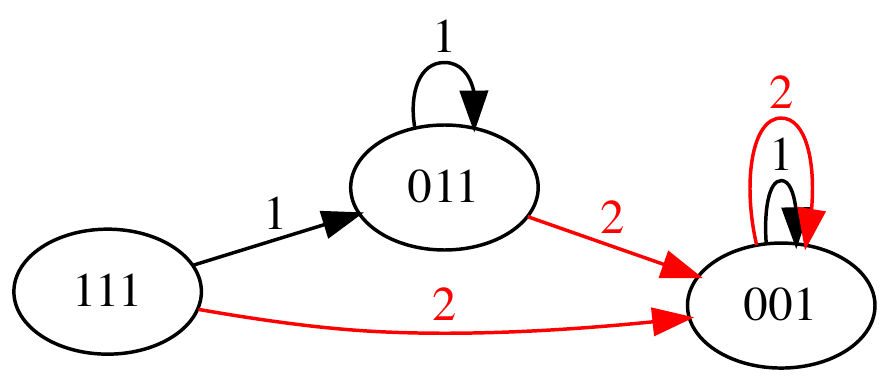}
\caption{\label{fig:park-mullins} 
Mullins effect. 
\textbf{(a)} Minimal schematic representation starting from a featureless state, as in Fig.~\ref{fig:minimal}. 
The state $\{2\}^*$ is incapable of having a memory of $1$ written in it. 
\textbf{(b)} Transition graph for the park bench model with $N=2$, $h_\text{init}=1$. 
A single memory is encoded where the grass height switches from $0$ to a plateau of $1$. 
}
\end{figure}

We now return to the case without noise to show that the park bench model can capture another distinct memory behavior. 
In particular, we note that a simpler form of memory occurs for $h_\text{init}=1$. 
Here there is no transient because a single cycle removes all the grass up to the visited bench; one might call this the ``scorched earth'' version of the model. 
Thus, the system remembers only the largest amplitude in its entire driving history. 
This is the same general behavior as the Mullins effect~\cite{Mullins48,Diani09,Schmoller13}, which occurs in polymer networks such as rubber under cyclic loading. 
There, the memory is indicated by a kink in the stress-strain curve at the largest stress that was previously applied to the sample. 
Figure~\ref{fig:park-mullins}a shows the minimal transition graph for the Mullins effect, which is equivalent to the park bench model with $N=2$, $h_\text{init}=1$, shown in Fig.~\ref{fig:park-mullins}b. 
One can easily construct the corresponding park bench graph for any $N$, which will have the same memory behavior.

\subsection{Return-point memory}\label{sec:RPM}

\begin{figure}[t]
\centering
\raisebox{0.7in}{\textbf{a)}} \includegraphics[scale=0.55]{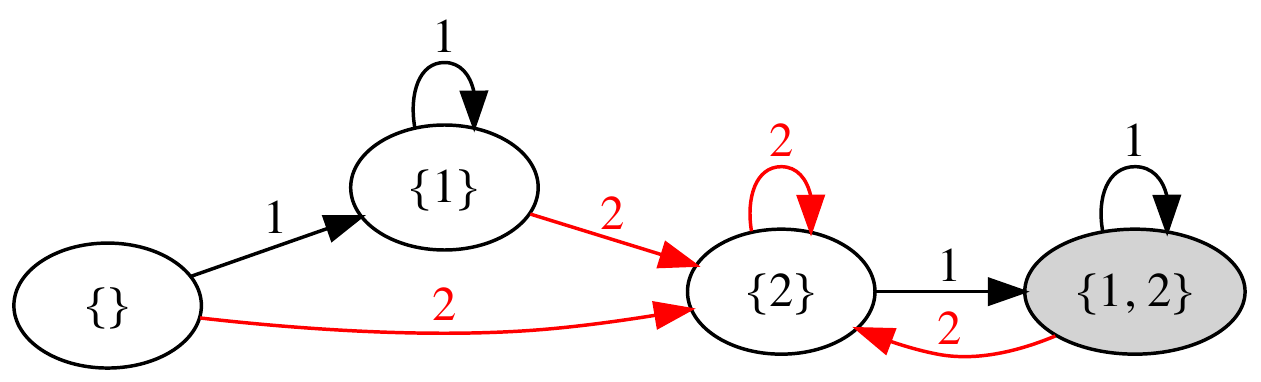} 
\hspace{0.25in}
\raisebox{0.7in}{\textbf{b)}} \includegraphics[scale=0.55]{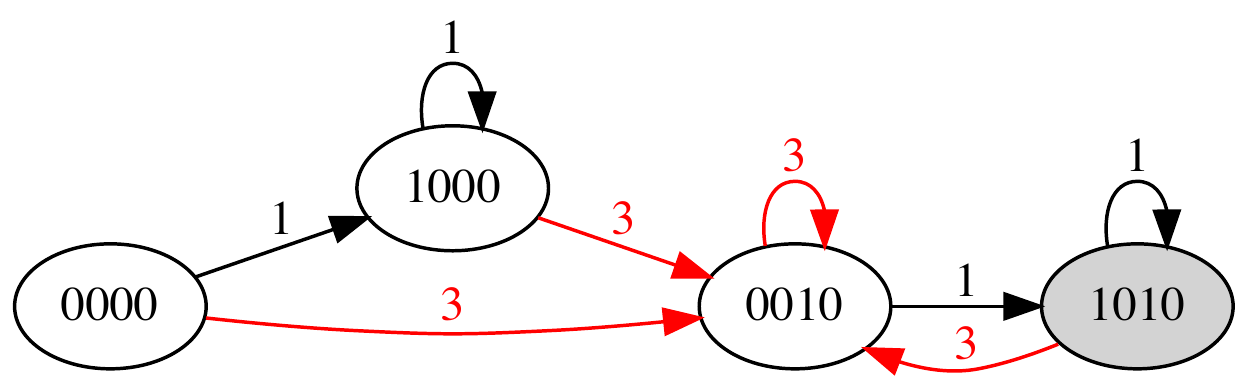}
\caption{\label{fig:rpm2cyc} 
Return-point memory (RPM). 
\textbf{(a)} Minimal schematic representation for the case of two driving amplitudes and starting from a featureless state, as in Fig.~\ref{fig:minimal}.
The order in which memories are written is important: to reach the multiple-memory state $\{1, 2\}$, the last driving must have amplitude $1$.
\textbf{(b)} The smallest realization of our ferromagnet model shows RPM.
The multiple memory state 1010 may only be reached by writing the small memory ($\gamma=1$) after the large one ($\gamma=3$). 
}
\end{figure}

To further demonstrate the generality of our approach of describing memory behaviors as properties of graphs of memory-encoding macrostates, we now develop a simple description of return-point memory 
(RPM). 
For cyclic driving, the key property of return-point memory can be described as follows.
Suppose a system is driven with an amplitude $\gamma_0$, thereby putting it in a state $s$. 
The system is then subjected to further driving cycles, all having amplitude less than or equal to $\gamma_0$. 
The system has return-point memory if a single cycle of amplitude $\gamma_0$ will then return the system to the exact same state, $s$; it \textit{remembers} this previous state. 
This generic behavior is observed in ferromagnets~\cite{Barker83,Sethna93} and many other non-equilibrium systems~\cite{Emmett47,Lilly93,Deutsch04,Ortin11}.
Because returning to $s$ is equivalent to wiping out all hysteresis since $s$ was last visited, the system's behavior can also change noticeably as $\gamma_0$ is surpassed, allowing the memory to be read out by a macroscopic observable such as magnetization.

\begin{figure*}[b]
\centering
\includegraphics[scale=0.52]{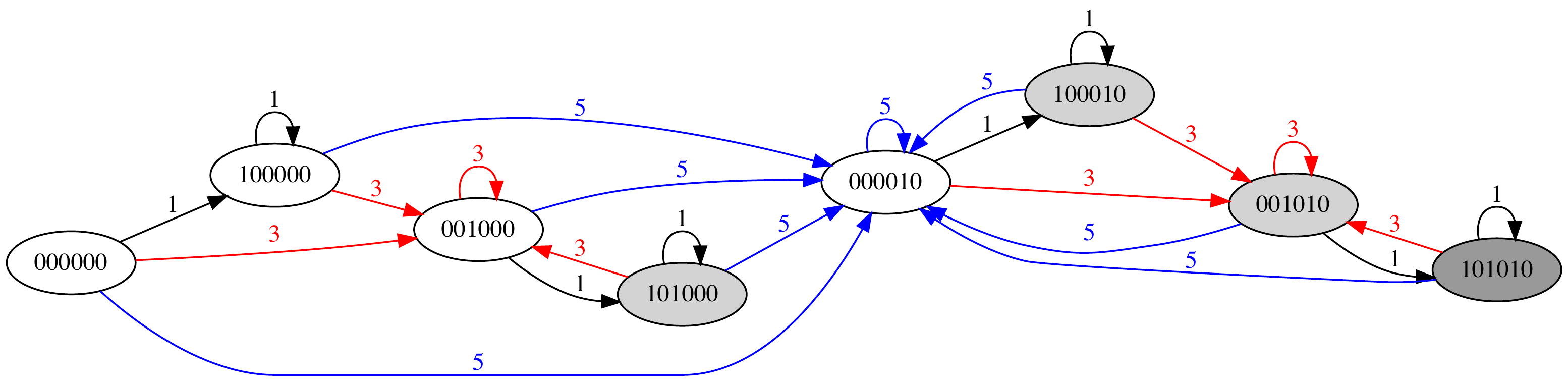}
\caption{\label{fig:rpm4cyc} 
Transition graph for our ferromagnet model with 6 hysterons. 
There are four different states that store two memories (101000, 101000, 100010, and 001010) and one state that stores three memories (101010). 
Memories are erased only by driving with a larger amplitude; the largest amplitude applied to the system over its entire history is thus always retained. 
}
\end{figure*}

\textit{Minimal graph of return-point memory ---}
Figure~\ref{fig:rpm2cyc}a shows a schematic depiction of the minimal set of states and transitions for RPM. 
The transition graph is strikingly similar to the graph of the Mullins effect in Fig.~\ref{fig:park-mullins}b, but with the addition of the multiple-memory state $\{1, 2\}$. 
This multiple-memory state can only be reached if the smaller amplitude is applied last. 

In general, graphs with RPM have the following distinct properties: 
(1) the ``maximal'' state (e.g., \{2\} in Fig.~\ref{fig:rpm2cyc}a) can be reached from any other state by applying the maximum allowed amplitude; 
(2) of all possible paths from the maximal state to any reachable state (here just \{1, 2\}) there is a unique path that does not involve erasure of a memory; and 
(3) there is no attractor with reduced memory, such as the $\{2\}^*$ state in MTM. 
Property 2 is a consequence of ``no-passing''~\cite{Middleton92} and expresses the importance of the order in which memories are added. 
Property 3 indicates that noise is not required to maintain the system's long-term capacity for memories. 

\textit{Simple model of return-point memory ---}
\label{sec:rpmmodel}
Figure~\ref{fig:rpm2cyc}b shows a transition graph where states are represented as binary strings of length $N$. 
Each transition represents a cycle with driving amplitude $H_0 < N$ in which the first $H-1$ digits are set to ``0'', and digit $H$ is set to ``1''. 
We restrict the amplitudes $H_0$ to odd integers less than $N$. 
A memory is indicated wherever the substring ``10'' appears. 
These rules reproduce the graph structure in Fig~\ref{fig:rpm2cyc}a. 

These rules for strings are motivated by a physical system: they arise from a simplified version of the Preisach model of a ferromagnet~\cite{Preisach35,Barker83}, which is a well-studied model for RPM. 
We use $N$ uncoupled spins (also called hysterons), indexed by $j=1,2, \ldots N$, which are driven by an external field, $H$. 
Each spin may be ``on'' with state $1$ or ``off'' with state $0$. 
In our model, the $j$th spin turns on at $H \ge H_\text{on}=j$ and off at $G \le H_\text{off}=-(j+1)$. 
We restrict ourselves to driving cycles following the sequence $H = 0 \to H_0 \to -H_0 \to 0$. 
(Note that $H$ plays the role of $\gamma$, but we use the symbol $H$ for familiarity.)  
States are denoted by a binary string of length $N$, indicating the state of each spin. 
As the field is ramped up from $0$ to $H_0$ it writes ``1'' on the string from left to right; as it is ramped down to $-H_0$ it writes ``0'' on the string from left to right. 
Thus, a cycle of amplitude $H_0$ overwrites the first $H_0-1$ characters in the string with ``0'' and writes a single ``1'' at position $H_0$. 

In a real ferromagnet, memories are read out by observing a discontinuity in the slope of a graph of magnetization (the average state of the spins) versus $H$. 
Here this occurs wherever the substring ``10'' appears in the string---a gap in the sequence of spin flips as $H$ is ramped up from 0. 
This method of readout requires that memories be separated, which is ensured by our restricting the driving to odd amplitudes that are less than $N$, so that 
all accessible states are sequences of the substrings ``00'' and ``10''. 

Figure~\ref{fig:rpm2cyc}b shows the smallest such model with RPM, $N=4$. 
Starting at the state $0000$, driving with amplitude $H_0=1$ leads to $1000$, which is a fixed point under repeated driving with $H_0=1$. 
Driving with $H_0=3$ leads to $0010$. 
From this state, $H_0=1$ adds a memory to the first position. 
In contrast to MTM the two memories cannot be written in any order; $H_0=1$ must be written last. 
Figure~\ref{fig:rpm4cyc} shows the transition graph for $N=6$. 
As in the smaller system, there is a unique path without erasure to any multiple-memory state. 
The graph also shows quite clearly that from any state, a single application of $H_0=5$ brings the system to the maximal state, $000010$, immediately erasing any smaller memories. 

To establish return point memory for arbitrary $N$, consider a cycle of amplitude $H_0$ that puts the system in state $s$. 
Suppose a sequence of $\{H_i\}$ with all $H_i < H_0$ is then applied. 
We must show that applying $H_0$ again returns the system to the state $s$. 
This may be seen by noting that the state $s$ starts with $H_0$ zeros. 
Each of the cycles of amplitude $H_i$  alters only the hysterons with indices $j<H_0$ (since $H_i < H_0$). 
A cycle of amplitude $H_0$ thus resets the first $H_0$ hysterons back to $0$.

\subsection{Diagnosing memory behavior in experiments}\label{sec:diagnose}

\begin{figure}[b]
\centering
\raisebox{0.85in}{\textbf{a)}} \includegraphics[scale=0.55]{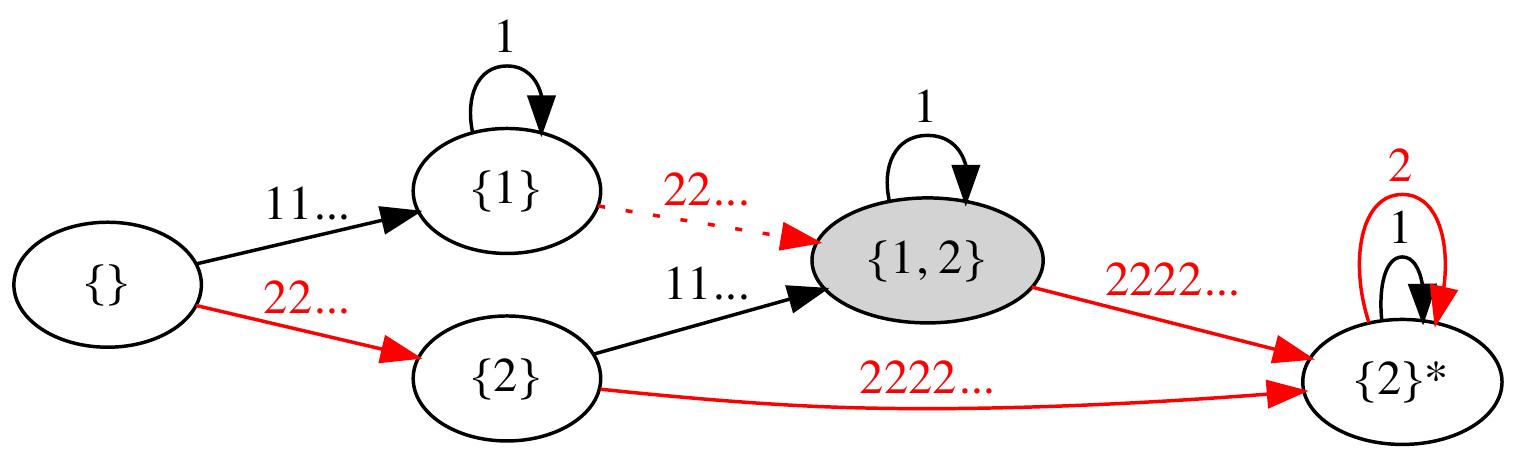} \\
\vspace{0.1in}
\raisebox{0.7in}{\textbf{b)}} \includegraphics[scale=0.54]{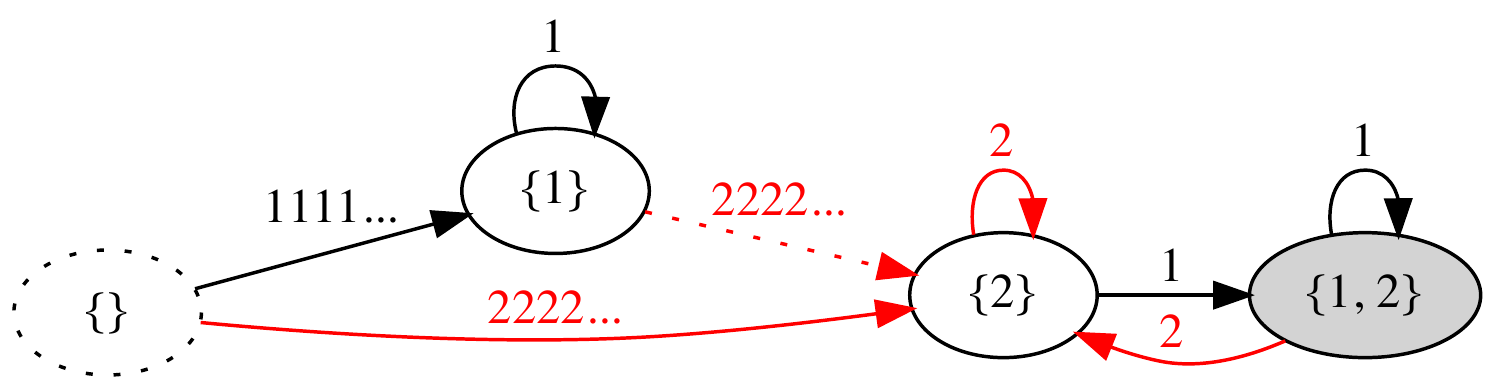}
\caption{\label{fig:experiment} 
States and transitions consistent with experimental data. Dotted states and transitions are present in models but not yet observed.
\textbf{(a)} Multiple transient memory in experiments on dilute non-Brownian suspensions, following structure of Fig.~\ref{fig:minimal}~\cite{Paulsen14}.
\textbf{(b)} Behavior similar to return-point memory, in experiments and simulations with two-dimensional amorphous solids, following structure of Fig.~\ref{fig:rpm2cyc}a~\cite{Adhikari18,Keim18}. The notation ``1111\dots'' and ``2222\dots'' represents evolution over many cycles of driving until a steady state is reached; ``22\dots'' represents multiple cycles that do not reach a steady state. In the case of MTM it may be possible to write a memory with only a single cycle, but experiments were not sensitive enough to detect this.
}
\end{figure}

Although we have illustrated our approach using simple models, we expect it should be useful for  experiments and dynamical simulations, by forming hypotheses and excluding possible memory behaviors. 
Dividing driving into cycles and reading out memories is a structured way to do this, and it lets us focus on memory-encoding macrostates, rather than the many microstates of a large system that occur within one cycle of driving, as in the work of Mungan and Terzi~\cite{Mungan18}.
For example, the minimal graph for MTM in Fig.~\ref{fig:minimal} possesses a state $\{2\}^*$, which has a memory of $2$ but with no capacity for writing a memory of $1$. 
In contrast, RPM does not have such a state; a smaller memory may always be written. 
A series of experiments on dilute suspensions recently established memory behavior consistent with MTM, which is represented in Fig.~\ref{fig:experiment}a~\cite{Paulsen14}. 
We point out that a small subset of those experiments---\emph{i.e.}, those establishing the existence of a state $\{2\}^*$---is enough to demonstrate that the memory behavior is distinct from RPM. 
Likewise, in experiments and simulations with amorphous solids, summarized in Fig.~\ref{fig:experiment}b, we can identify an analog of each state in Fig.~\ref{fig:rpm2cyc}a, show the \emph{absence} of an absorbing state, and demonstrate that memory content depends on which amplitude was applied last, suggesting a behavior similar to RPM~\cite{Adhikari18,Keim18}. 

Figure.~\ref{fig:experiment} also demonstrates the value of transition graphs in organizing experimental results. By enumerating a complete set of transitions, one can identify which transitions have been demonstrated experimentally. For instance, experiments on dilute suspensions have not yet shown the existence of the $\{1\} \to \{1, 2\}$ transition in Fig.~\ref{fig:experiment}a, despite establishing all other characteristics of multiple transient memories~\cite{Paulsen14}.

In systems with significant transients, our general approach becomes more cumbersome due to a possibly large number of states. 
In this case we may consider the effective transitions that result from either driving the system with a large number of cycles much greater than the known transient length, denoted as ``1111\dots'' or ``2222\dots'' in Fig.~\ref{fig:experiment}, or from driving with an intermediate number of cycles less than the transient length, denoted as ``11\dots'' and ``22\dots''. 

We note that developing a reliable readout is essential for these applications.
An example of a readout protocol is to drive the system with a series of increasing amplitudes and monitor its response as a function of amplitude~\cite{Fiocco14,Paulsen14,Keim18}. 
In this case, one must be careful to ensure that the readout process does not introduce or erase memories before they are observed. 
One way to test this property is to compare the results of a ``sequential'' readout involving a series of cycles at different amplitudes, with a ``parallel'' readout that takes many identically-prepared systems (or copies of a single simulated system) and then drives each with a different amplitude for one cycle~\cite{Sood18,Adhikari18}. 
Once a method for readout is established, mapping some or all of the states reached by cyclic driving can be a straightforward yet powerful diagnostic test. 

One may also construct even simpler tests that do not require a careful memory readout. 
For instance, if one can compare states of the system observed after each cycle of driving, \emph{e.g.} by simple image subtraction, one can identify distinct states and map their transitions without knowing their memory content. 
In dilute suspensions, when many cycles of a given amplitude are applied, further driving at a smaller amplitude does not change the state (away from $\{2\}^*$)~\cite{Paulsen14}. 
This by itself rules out RPM. 
The same test in amorphous solids \emph{does} change the state (to $\{1, 2\}$)~\cite{Keim18}.

\section{Discussion}

This work establishes a simple graph structure as a common language for comparing memories across multiple systems. 
This may help to sort through the growing body of work on cyclic memory formation and self-organization. 
This includes the recent findings that multiple transient memory may occur in seemingly disparate models and physical systems~\cite{Coppersmith97,Povinelli99,Keim11,Paulsen14,Adhikari18}, but also some less-understood examples, such as the evolution of bandgaps in a 1D array of particles driven by acoustic waves~\cite{Bachelard17}, and cyclic memories observed in glassy systems like amorphous solids~\cite{Fiocco14,Fiocco15,Adhikari18,Keim18}. Moreover this approach could help to identify memory in subgraphs that are embedded in a larger set of states, similar to the more detailed description of return-point memory developed by Mungan and Terzi~\cite{Mungan18}. Finally, it lets us imagine new, as-yet undiscovered cyclic memory behaviors and consider how they might be identified.

We also demonstrated the minimal set of states required for multiple transient memory, and we described a simple physically-motivated model that produces this behavior. 
The park bench model of MTM can store multiple pieces of information (i.e., locations of jumps in the grass height) in transient states, but it forgets all but the largest repeated excursion in the steady state. 
This is somewhat remarkable as the system has a dearth of complexity: there is only a small, enumerable set of states, no disorder, and the evolution is determined by the sequence of inputs with no stochastic element. 

When noise is added to the park bench model, all memories were stabilized at long times, consistent with other systems with MTM and noise (MTMN)~\cite{Coppersmith97, Povinelli99, Keim11, Keim13, Paulsen14}.
By considering the size of noise-induced fluctuations in a steady state under repeated driving, we demonstrated a route to assessing the memory capacity of MTMN. This led to an analytic 
estimate for the memory capacity of the noisy park bench model, and a way to model the results of arbitrary readout protocols. 
Memory capacity has received considerably more attention in models of associative memory \cite{Hopfield82,Parisi86}, and in more realistic models of biological neural networks \cite{amit_1989,Fusi2017}. 
Similar to these more-complex neural networks, the noisy park bench model also displays plasticity: We find that after reaching a steady state with one driving amplitude, we can switch to another amplitude and form a new memory of that value instead. 
(This outcome of MTMN was previously found in a model of cyclically-sheared suspensions with noise \cite{Keim13}.)

The park bench model also demonstrates that criticality is not required for MTM. 
In some other forms of memory such as aging and rejuvenation in glasses~\cite{Jonason98,Yardimci03,Yang17}, 
multiple memories may exist simultaneously because the system has many relaxation processes across a range of length- and timescales. 
Proximity to a critical point is a natural way to get this wide range of scales, suggesting a link between multiple memory formation and criticality. 
Indeed, sheared non-Brownian suspensions and charge-density wave conductors both feature critical transitions in their dynamics---a depinning transition of the charge-density wave \cite{Fisher85, Coppersmith87}, and an irreversibility transition of the sheared suspension with diverging time- and length-scales \cite{Corte08, Keim11, Keim13}. 
But this is just one strategy for avoiding interference of multiple memories; a simpler strategy is for the driving to select a unique scale directly, as occurs in the park bench model and our simple ferromagnet model. 

Recent studies of memory formation in sheared non-Brownian suspensions~\cite{Keim11,Keim13,Paulsen14}, amorphous solids
~\cite{Fiocco14,Fiocco15,Adhikari18,Keim18}, frustrated spin systems~\cite{Fiocco15}, and charge-density waves~\cite{Coppersmith97,Povinelli99}, have raised the tantalizing possibility that systems with the same memory behavior may share deeper aspects of their physics, such as a critical transition. 
The existence of a physically-motivated model of multiple transient memory that has neither criticality nor nonlinear diffusion suggests that this idea should be pursued with caution. 
On the other hand, it shows that an extremely simple model can elucidate underlying mechanisms for memory behaviors. 
A similar approach has been illuminating in the study of aging and rejuvenation in glasses, where a simple algorithm that sorts a short list of numbers was found to capture a non-trivial set of memory behaviors~\cite{Zou10}.

\bigskip

\medskip\noindent\textbf{Data accessibility.} Data for Fig.~\ref{fig:noisy}a, and Python code to generate all simulation results, are available in the electronic supplementary material.

\medskip\noindent\textbf{Competing interests.} The authors declare no competing interests. 

\medskip\noindent\textbf{Authors' contributions.} J.D.P. and N.C.K. conceived the study and analyzed the models. N.C.K. performed simulations. J.D.P. and N.C.K. interpreted the results and wrote the paper. 

\medskip\noindent\textbf{Acknowledgements.} This work was initiated at the Winter 2018 KITP program, ``Memory Formation in Matter.'' 
We are grateful to Sidney Nagel for inventing the park bench model, and Muhittin Mungan for proposing and encouraging the use of graphs. 
We also thank Srikanth Sastry, Tom Witten, and other participants in the program. 

\medskip\noindent\textbf{Funding statement.} J.D.P.\ gratefully acknowledges the Donors of the American Chemical Society Petroleum Research Fund for partial support of this research. 
This research was also supported by the National Science Foundation under Grant No.\ PHY-1748958 to the KITP, and Grant No.\ DMR-1708870 to N.C.K. 

\medskip\noindent\textbf{Ethics statement.} This work involves data obtained only from computer simulations.

\bibliography{references}

\begin{thebibliography}{43}%
\makeatletter
\providecommand \@ifxundefined [1]{%
 \@ifx{#1\undefined}
}%
\providecommand \@ifnum [1]{%
 \ifnum #1\expandafter \@firstoftwo
 \else \expandafter \@secondoftwo
 \fi
}%
\providecommand \@ifx [1]{%
 \ifx #1\expandafter \@firstoftwo
 \else \expandafter \@secondoftwo
 \fi
}%
\providecommand \natexlab [1]{#1}%
\providecommand \enquote  [1]{``#1''}%
\providecommand \bibnamefont  [1]{#1}%
\providecommand \bibfnamefont [1]{#1}%
\providecommand \citenamefont [1]{#1}%
\providecommand \href@noop [0]{\@secondoftwo}%
\providecommand \href [0]{\begingroup \@sanitize@url \@href}%
\providecommand \@href[1]{\@@startlink{#1}\@@href}%
\providecommand \@@href[1]{\endgroup#1\@@endlink}%
\providecommand \@sanitize@url [0]{\catcode `\\12\catcode `\$12\catcode
  `\&12\catcode `\#12\catcode `\^12\catcode `\_12\catcode `\%12\relax}%
\providecommand \@@startlink[1]{}%
\providecommand \@@endlink[0]{}%
\providecommand \url  [0]{\begingroup\@sanitize@url \@url }%
\providecommand \@url [1]{\endgroup\@href {#1}{\urlprefix }}%
\providecommand \urlprefix  [0]{URL }%
\providecommand \Eprint [0]{\href }%
\providecommand \doibase [0]{http://dx.doi.org/}%
\providecommand \selectlanguage [0]{\@gobble}%
\providecommand \bibinfo  [0]{\@secondoftwo}%
\providecommand \bibfield  [0]{\@secondoftwo}%
\providecommand \translation [1]{[#1]}%
\providecommand \BibitemOpen [0]{}%
\providecommand \bibitemStop [0]{}%
\providecommand \bibitemNoStop [0]{.\EOS\space}%
\providecommand \EOS [0]{\spacefactor3000\relax}%
\providecommand \BibitemShut  [1]{\csname bibitem#1\endcsname}%
\let\auto@bib@innerbib\@empty
\bibitem [{\citenamefont {Mullins}(1948)}]{Mullins48}%
  \BibitemOpen
  \bibfield  {author} {\bibinfo {author} {\bibfnamefont {L}~\bibnamefont
  {Mullins}},\ }\bibfield  {title} {\enquote {\bibinfo {title} {Effect of
  stretching on the properties of rubber},}\ }\href@noop {} {\bibfield
  {journal} {\bibinfo  {journal} {Rubber Chemistry and Technology}\ }\textbf
  {\bibinfo {volume} {21}},\ \bibinfo {pages} {281--300} (\bibinfo {year}
  {1948})}\BibitemShut {NoStop}%
\bibitem [{\citenamefont {Kurita}\ and\ \citenamefont
  {Fujii}(1979)}]{Kurita79}%
  \BibitemOpen
  \bibfield  {author} {\bibinfo {author} {\bibfnamefont {Kei}\ \bibnamefont
  {Kurita}}\ and\ \bibinfo {author} {\bibfnamefont {Naoyuki}\ \bibnamefont
  {Fujii}},\ }\bibfield  {title} {\enquote {\bibinfo {title} {Stress memory of
  crystalline rocks in acoustic emission},}\ }\href@noop {} {\bibfield
  {journal} {\bibinfo  {journal} {Geophysical Research Letters}\ }\textbf
  {\bibinfo {volume} {6}},\ \bibinfo {pages} {9--12} (\bibinfo {year}
  {1979})}\BibitemShut {NoStop}%
\bibitem [{\citenamefont {Schmoller}\ and\ \citenamefont
  {Bausch}(2013)}]{Schmoller13}%
  \BibitemOpen
  \bibfield  {author} {\bibinfo {author} {\bibfnamefont {Kurt~M}\ \bibnamefont
  {Schmoller}}\ and\ \bibinfo {author} {\bibfnamefont {Andreas~R}\ \bibnamefont
  {Bausch}},\ }\bibfield  {title} {\enquote {\bibinfo {title} {Similar
  nonlinear mechanical responses in hard and soft materials},}\ }\href@noop {}
  {\bibfield  {journal} {\bibinfo  {journal} {Nature materials}\ }\textbf
  {\bibinfo {volume} {12}},\ \bibinfo {pages} {278} (\bibinfo {year}
  {2013})}\BibitemShut {NoStop}%
\bibitem [{\citenamefont {Jonason}\ \emph {et~al.}(1998)\citenamefont
  {Jonason}, \citenamefont {Vincent}, \citenamefont {Hammann}, \citenamefont
  {Bouchaud},\ and\ \citenamefont {Nordblad}}]{Jonason98}%
  \BibitemOpen
  \bibfield  {author} {\bibinfo {author} {\bibfnamefont {K}~\bibnamefont
  {Jonason}}, \bibinfo {author} {\bibfnamefont {E}~\bibnamefont {Vincent}},
  \bibinfo {author} {\bibfnamefont {J}~\bibnamefont {Hammann}}, \bibinfo
  {author} {\bibfnamefont {JP}~\bibnamefont {Bouchaud}}, \ and\ \bibinfo
  {author} {\bibfnamefont {P}~\bibnamefont {Nordblad}},\ }\bibfield  {title}
  {\enquote {\bibinfo {title} {Memory and chaos effects in spin glasses},}\
  }\href@noop {} {\bibfield  {journal} {\bibinfo  {journal} {Physical Review
  Letters}\ }\textbf {\bibinfo {volume} {81}},\ \bibinfo {pages} {3243}
  (\bibinfo {year} {1998})}\BibitemShut {NoStop}%
\bibitem [{\citenamefont {Zou}\ and\ \citenamefont {Nagel}(2010)}]{Zou10}%
  \BibitemOpen
  \bibfield  {author} {\bibinfo {author} {\bibfnamefont {Ling-Nan}\
  \bibnamefont {Zou}}\ and\ \bibinfo {author} {\bibfnamefont {Sidney~R}\
  \bibnamefont {Nagel}},\ }\bibfield  {title} {\enquote {\bibinfo {title}
  {Glassy dynamics in thermally activated list sorting},}\ }\href@noop {}
  {\bibfield  {journal} {\bibinfo  {journal} {Physical review letters}\
  }\textbf {\bibinfo {volume} {104}},\ \bibinfo {pages} {257201} (\bibinfo
  {year} {2010})}\BibitemShut {NoStop}%
\bibitem [{\citenamefont {Gilbert}\ \emph {et~al.}(2015)\citenamefont
  {Gilbert}, \citenamefont {Chern}, \citenamefont {Fore}, \citenamefont {Lao},
  \citenamefont {Zhang}, \citenamefont {Nisoli},\ and\ \citenamefont
  {Schiffer}}]{Gilbert15}%
  \BibitemOpen
  \bibfield  {author} {\bibinfo {author} {\bibfnamefont {Ian}\ \bibnamefont
  {Gilbert}}, \bibinfo {author} {\bibfnamefont {Gia-Wei}\ \bibnamefont
  {Chern}}, \bibinfo {author} {\bibfnamefont {Bryce}\ \bibnamefont {Fore}},
  \bibinfo {author} {\bibfnamefont {Yuyang}\ \bibnamefont {Lao}}, \bibinfo
  {author} {\bibfnamefont {Sheng}\ \bibnamefont {Zhang}}, \bibinfo {author}
  {\bibfnamefont {Cristiano}\ \bibnamefont {Nisoli}}, \ and\ \bibinfo {author}
  {\bibfnamefont {Peter}\ \bibnamefont {Schiffer}},\ }\bibfield  {title}
  {\enquote {\bibinfo {title} {{Direct visualization of memory effects in
  artificial spin ice}},}\ }\href@noop {} {\bibfield  {journal} {\bibinfo
  {journal} {Phys. Rev. B}\ }\textbf {\bibinfo {volume} {92}},\ \bibinfo
  {pages} {104417} (\bibinfo {year} {2015})}\BibitemShut {NoStop}%
\bibitem [{\citenamefont {Yang}\ and\ \citenamefont
  {Middleton}(2017)}]{Yang17}%
  \BibitemOpen
  \bibfield  {author} {\bibinfo {author} {\bibfnamefont {Jie}\ \bibnamefont
  {Yang}}\ and\ \bibinfo {author} {\bibfnamefont {A~Alan}\ \bibnamefont
  {Middleton}},\ }\bibfield  {title} {\enquote {\bibinfo {title} {Configuration
  memory in patchwork dynamics for low-dimensional spin glasses},}\ }\href@noop
  {} {\bibfield  {journal} {\bibinfo  {journal} {Physical Review B}\ }\textbf
  {\bibinfo {volume} {96}},\ \bibinfo {pages} {214208} (\bibinfo {year}
  {2017})}\BibitemShut {NoStop}%
\bibitem [{\citenamefont {Matan}\ \emph {et~al.}(2002)\citenamefont {Matan},
  \citenamefont {Williams}, \citenamefont {Witten},\ and\ \citenamefont
  {Nagel}}]{Matan02}%
  \BibitemOpen
  \bibfield  {author} {\bibinfo {author} {\bibfnamefont {Kittiwit}\
  \bibnamefont {Matan}}, \bibinfo {author} {\bibfnamefont {Rachel~B}\
  \bibnamefont {Williams}}, \bibinfo {author} {\bibfnamefont {Thomas~A}\
  \bibnamefont {Witten}}, \ and\ \bibinfo {author} {\bibfnamefont {Sidney~R}\
  \bibnamefont {Nagel}},\ }\bibfield  {title} {\enquote {\bibinfo {title}
  {Crumpling a thin sheet},}\ }\href@noop {} {\bibfield  {journal} {\bibinfo
  {journal} {Physical Review Letters}\ }\textbf {\bibinfo {volume} {88}},\
  \bibinfo {pages} {076101} (\bibinfo {year} {2002})}\BibitemShut {NoStop}%
\bibitem [{\citenamefont {Lahini}\ \emph {et~al.}(2017)\citenamefont {Lahini},
  \citenamefont {Gottesman}, \citenamefont {Amir},\ and\ \citenamefont
  {Rubinstein}}]{Lahini17}%
  \BibitemOpen
  \bibfield  {author} {\bibinfo {author} {\bibfnamefont {Yoav}\ \bibnamefont
  {Lahini}}, \bibinfo {author} {\bibfnamefont {Omer}\ \bibnamefont
  {Gottesman}}, \bibinfo {author} {\bibfnamefont {Ariel}\ \bibnamefont {Amir}},
  \ and\ \bibinfo {author} {\bibfnamefont {Shmuel~M}\ \bibnamefont
  {Rubinstein}},\ }\bibfield  {title} {\enquote {\bibinfo {title} {Nonmonotonic
  aging and memory retention in disordered mechanical systems},}\ }\href@noop
  {} {\bibfield  {journal} {\bibinfo  {journal} {Physical review letters}\
  }\textbf {\bibinfo {volume} {118}},\ \bibinfo {pages} {085501} (\bibinfo
  {year} {2017})}\BibitemShut {NoStop}%
\bibitem [{\citenamefont {Keim}\ \emph {et~al.}(2019)\citenamefont {Keim},
  \citenamefont {Paulsen}, \citenamefont {Zeravcic}, \citenamefont {Sastry},\
  and\ \citenamefont {Nagel}}]{Keim18review}%
  \BibitemOpen
  \bibfield  {author} {\bibinfo {author} {\bibfnamefont {Nathan~C}\
  \bibnamefont {Keim}}, \bibinfo {author} {\bibfnamefont {Joseph}\ \bibnamefont
  {Paulsen}}, \bibinfo {author} {\bibfnamefont {Zorana}\ \bibnamefont
  {Zeravcic}}, \bibinfo {author} {\bibfnamefont {Srikanth}\ \bibnamefont
  {Sastry}}, \ and\ \bibinfo {author} {\bibfnamefont {Sidney~R}\ \bibnamefont
  {Nagel}},\ }\bibfield  {title} {\enquote {\bibinfo {title} {Memory formation
  in matter},}\ }\href@noop {} {\bibfield  {journal} {\bibinfo  {journal}
  {Rev.\ Mod.\ Phys., in press. arXiv:1810.08587}\ } (\bibinfo {year}
  {2019})}\BibitemShut {NoStop}%
\bibitem [{\citenamefont {Toiya}\ \emph {et~al.}(2004)\citenamefont {Toiya},
  \citenamefont {Stambaugh},\ and\ \citenamefont {Losert}}]{Toiya04}%
  \BibitemOpen
  \bibfield  {author} {\bibinfo {author} {\bibfnamefont {Masahiro}\
  \bibnamefont {Toiya}}, \bibinfo {author} {\bibfnamefont {Justin}\
  \bibnamefont {Stambaugh}}, \ and\ \bibinfo {author} {\bibfnamefont
  {Wolfgang}\ \bibnamefont {Losert}},\ }\bibfield  {title} {\enquote {\bibinfo
  {title} {{Transient and Oscillatory Granular Shear Flow}},}\ }\href@noop {}
  {\bibfield  {journal} {\bibinfo  {journal} {Phys. Rev. Lett.}\ }\textbf
  {\bibinfo {volume} {93}},\ \bibinfo {pages} {088001} (\bibinfo {year}
  {2004})}\BibitemShut {NoStop}%
\bibitem [{\citenamefont {Royer}\ and\ \citenamefont
  {Chaikin}(2015)}]{Royer15}%
  \BibitemOpen
  \bibfield  {author} {\bibinfo {author} {\bibfnamefont {John~R}\ \bibnamefont
  {Royer}}\ and\ \bibinfo {author} {\bibfnamefont {Paul~M}\ \bibnamefont
  {Chaikin}},\ }\bibfield  {title} {\enquote {\bibinfo {title} {Precisely
  cyclic sand: Self-organization of periodically sheared frictional grains},}\
  }\href@noop {} {\bibfield  {journal} {\bibinfo  {journal} {Proceedings of the
  National Academy of Sciences}\ }\textbf {\bibinfo {volume} {112}},\ \bibinfo
  {pages} {49--53} (\bibinfo {year} {2015})}\BibitemShut {NoStop}%
\bibitem [{\citenamefont {Fiocco}\ \emph {et~al.}(2014)\citenamefont {Fiocco},
  \citenamefont {Foffi},\ and\ \citenamefont {Sastry}}]{Fiocco14}%
  \BibitemOpen
  \bibfield  {author} {\bibinfo {author} {\bibfnamefont {Davide}\ \bibnamefont
  {Fiocco}}, \bibinfo {author} {\bibfnamefont {Giuseppe}\ \bibnamefont
  {Foffi}}, \ and\ \bibinfo {author} {\bibfnamefont {Srikanth}\ \bibnamefont
  {Sastry}},\ }\bibfield  {title} {\enquote {\bibinfo {title} {{Encoding of
  Memory in Sheared Amorphous Solids}},}\ }\href@noop {} {\bibfield  {journal}
  {\bibinfo  {journal} {Phys. Rev. Lett.}\ }\textbf {\bibinfo {volume} {112}},\
  \bibinfo {pages} {025702} (\bibinfo {year} {2014})}\BibitemShut {NoStop}%
\bibitem [{\citenamefont {Keim}\ and\ \citenamefont {Arratia}(2014)}]{Keim14}%
  \BibitemOpen
  \bibfield  {author} {\bibinfo {author} {\bibfnamefont {Nathan~C}\
  \bibnamefont {Keim}}\ and\ \bibinfo {author} {\bibfnamefont {Paulo~E}\
  \bibnamefont {Arratia}},\ }\bibfield  {title} {\enquote {\bibinfo {title}
  {{Mechanical and Microscopic Properties of the Reversible Plastic Regime in a
  2D Jammed Material}},}\ }\href@noop {} {\bibfield  {journal} {\bibinfo
  {journal} {Phys. Rev. Lett.}\ }\textbf {\bibinfo {volume} {112}},\ \bibinfo
  {pages} {028302} (\bibinfo {year} {2014})}\BibitemShut {NoStop}%
\bibitem [{\citenamefont {Barker}\ \emph {et~al.}(1983)\citenamefont {Barker},
  \citenamefont {Schreiber}, \citenamefont {Huth},\ and\ \citenamefont
  {Everett}}]{Barker83}%
  \BibitemOpen
  \bibfield  {author} {\bibinfo {author} {\bibfnamefont {J}~\bibnamefont
  {Barker}}, \bibinfo {author} {\bibfnamefont {D}~\bibnamefont {Schreiber}},
  \bibinfo {author} {\bibfnamefont {BG}~\bibnamefont {Huth}}, \ and\ \bibinfo
  {author} {\bibfnamefont {D~H}\ \bibnamefont {Everett}},\ }\bibfield  {title}
  {\enquote {\bibinfo {title} {{Magnetic hysteresis and minor loops: Models and
  experiments}},}\ }\href@noop {} {\bibfield  {journal} {\bibinfo  {journal}
  {Proc. R. Soc. Lond. A}\ }\textbf {\bibinfo {volume} {386}},\ \bibinfo
  {pages} {251--261} (\bibinfo {year} {1983})}\BibitemShut {NoStop}%
\bibitem [{\citenamefont {Sethna}\ \emph {et~al.}(1993)\citenamefont {Sethna},
  \citenamefont {Dahmen}, \citenamefont {Kartha}, \citenamefont {Krumhansl},
  \citenamefont {Roberts},\ and\ \citenamefont {Shore}}]{Sethna93}%
  \BibitemOpen
  \bibfield  {author} {\bibinfo {author} {\bibfnamefont {James~P}\ \bibnamefont
  {Sethna}}, \bibinfo {author} {\bibfnamefont {Karin}\ \bibnamefont {Dahmen}},
  \bibinfo {author} {\bibfnamefont {Sivan}\ \bibnamefont {Kartha}}, \bibinfo
  {author} {\bibfnamefont {James~A}\ \bibnamefont {Krumhansl}}, \bibinfo
  {author} {\bibfnamefont {Bruce~W}\ \bibnamefont {Roberts}}, \ and\ \bibinfo
  {author} {\bibfnamefont {Joel~D}\ \bibnamefont {Shore}},\ }\bibfield  {title}
  {\enquote {\bibinfo {title} {{Hysteresis and hierarchies: Dynamics of
  disorder-driven first-order phase transformations}},}\ }\href@noop {}
  {\bibfield  {journal} {\bibinfo  {journal} {Phys. Rev. Lett.}\ }\textbf
  {\bibinfo {volume} {70}},\ \bibinfo {pages} {3347} (\bibinfo {year}
  {1993})}\BibitemShut {NoStop}%
\bibitem [{\citenamefont {Mungan}\ and\ \citenamefont
  {Terzi}(2018)}]{Mungan18}%
  \BibitemOpen
  \bibfield  {author} {\bibinfo {author} {\bibfnamefont {Muhittin}\
  \bibnamefont {Mungan}}\ and\ \bibinfo {author} {\bibfnamefont {M~Mert}\
  \bibnamefont {Terzi}},\ }\bibfield  {title} {\enquote {\bibinfo {title} {The
  structure of state transition graphs in hysteresis models with return point
  memory. i. general theory},}\ }\href@noop {} {\bibfield  {journal} {\bibinfo
  {journal} {arXiv:1802.03096}\ } (\bibinfo {year} {2018})}\BibitemShut
  {NoStop}%
\bibitem [{\citenamefont {Fiocco}\ \emph {et~al.}(2015)\citenamefont {Fiocco},
  \citenamefont {Foffi},\ and\ \citenamefont {Sastry}}]{Fiocco15}%
  \BibitemOpen
  \bibfield  {author} {\bibinfo {author} {\bibfnamefont {Davide}\ \bibnamefont
  {Fiocco}}, \bibinfo {author} {\bibfnamefont {Giuseppe}\ \bibnamefont
  {Foffi}}, \ and\ \bibinfo {author} {\bibfnamefont {Srikanth}\ \bibnamefont
  {Sastry}},\ }\bibfield  {title} {\enquote {\bibinfo {title} {{Memory effects
  in schematic models of glasses subjected to oscillatory deformation}},}\
  }\href@noop {} {\bibfield  {journal} {\bibinfo  {journal} {J. Phys: Cond.
  Matter}\ }\textbf {\bibinfo {volume} {27}},\ \bibinfo {pages} {194130}
  (\bibinfo {year} {2015})}\BibitemShut {NoStop}%
\bibitem [{\citenamefont {Coppersmith}\ \emph {et~al.}(1997)\citenamefont
  {Coppersmith}, \citenamefont {Jones}, \citenamefont {Kadanoff}, \citenamefont
  {Levine}, \citenamefont {McCarten}, \citenamefont {Nagel}, \citenamefont
  {Venkataramani},\ and\ \citenamefont {Wu}}]{Coppersmith97}%
  \BibitemOpen
  \bibfield  {author} {\bibinfo {author} {\bibfnamefont {S~N}\ \bibnamefont
  {Coppersmith}}, \bibinfo {author} {\bibfnamefont {T~C}\ \bibnamefont
  {Jones}}, \bibinfo {author} {\bibfnamefont {L~P}\ \bibnamefont {Kadanoff}},
  \bibinfo {author} {\bibfnamefont {A}~\bibnamefont {Levine}}, \bibinfo
  {author} {\bibfnamefont {J~P}\ \bibnamefont {McCarten}}, \bibinfo {author}
  {\bibfnamefont {S~R}\ \bibnamefont {Nagel}}, \bibinfo {author} {\bibfnamefont
  {S~C}\ \bibnamefont {Venkataramani}}, \ and\ \bibinfo {author} {\bibfnamefont
  {Xinlei}\ \bibnamefont {Wu}},\ }\bibfield  {title} {\enquote {\bibinfo
  {title} {{Self-Organized Short-Term Memories}},}\ }\href@noop {} {\bibfield
  {journal} {\bibinfo  {journal} {Phys. Rev. Lett.}\ }\textbf {\bibinfo
  {volume} {78}},\ \bibinfo {pages} {3983--3986} (\bibinfo {year}
  {1997})}\BibitemShut {NoStop}%
\bibitem [{\citenamefont {Povinelli}\ \emph {et~al.}(1999)\citenamefont
  {Povinelli}, \citenamefont {Coppersmith}, \citenamefont {Kadanoff},
  \citenamefont {Nagel},\ and\ \citenamefont {Venkataramani}}]{Povinelli99}%
  \BibitemOpen
  \bibfield  {author} {\bibinfo {author} {\bibfnamefont {M~L}\ \bibnamefont
  {Povinelli}}, \bibinfo {author} {\bibfnamefont {S~N}\ \bibnamefont
  {Coppersmith}}, \bibinfo {author} {\bibfnamefont {L~P}\ \bibnamefont
  {Kadanoff}}, \bibinfo {author} {\bibfnamefont {S~R}\ \bibnamefont {Nagel}}, \
  and\ \bibinfo {author} {\bibfnamefont {S~C}\ \bibnamefont {Venkataramani}},\
  }\bibfield  {title} {\enquote {\bibinfo {title} {{Noise stabilization of
  self-organized memories}},}\ }\href@noop {} {\bibfield  {journal} {\bibinfo
  {journal} {Phys. Rev. E}\ }\textbf {\bibinfo {volume} {59}},\ \bibinfo
  {pages} {4970--4982} (\bibinfo {year} {1999})}\BibitemShut {NoStop}%
\bibitem [{\citenamefont {Keim}\ and\ \citenamefont {Nagel}(2011)}]{Keim11}%
  \BibitemOpen
  \bibfield  {author} {\bibinfo {author} {\bibfnamefont {Nathan~C}\
  \bibnamefont {Keim}}\ and\ \bibinfo {author} {\bibfnamefont {Sidney~R}\
  \bibnamefont {Nagel}},\ }\bibfield  {title} {\enquote {\bibinfo {title}
  {{Generic Transient Memory Formation in Disordered Systems with Noise}},}\
  }\href@noop {} {\bibfield  {journal} {\bibinfo  {journal} {Phys. Rev. Lett.}\
  }\textbf {\bibinfo {volume} {107}},\ \bibinfo {pages} {010603} (\bibinfo
  {year} {2011})}\BibitemShut {NoStop}%
\bibitem [{\citenamefont {Keim}\ \emph {et~al.}(2013)\citenamefont {Keim},
  \citenamefont {Paulsen},\ and\ \citenamefont {Nagel}}]{Keim13}%
  \BibitemOpen
  \bibfield  {author} {\bibinfo {author} {\bibfnamefont {Nathan~C}\
  \bibnamefont {Keim}}, \bibinfo {author} {\bibfnamefont {Joseph~D}\
  \bibnamefont {Paulsen}}, \ and\ \bibinfo {author} {\bibfnamefont {Sidney~R}\
  \bibnamefont {Nagel}},\ }\bibfield  {title} {\enquote {\bibinfo {title}
  {{Multiple transient memories in sheared suspensions: Robustness, structure,
  and routes to plasticity}},}\ }\href@noop {} {\bibfield  {journal} {\bibinfo
  {journal} {Phys. Rev. E}\ }\textbf {\bibinfo {volume} {88}},\ \bibinfo
  {pages} {032306} (\bibinfo {year} {2013})}\BibitemShut {NoStop}%
\bibitem [{\citenamefont {Paulsen}\ \emph {et~al.}(2014)\citenamefont
  {Paulsen}, \citenamefont {Keim},\ and\ \citenamefont {Nagel}}]{Paulsen14}%
  \BibitemOpen
  \bibfield  {author} {\bibinfo {author} {\bibfnamefont {Joseph~D}\
  \bibnamefont {Paulsen}}, \bibinfo {author} {\bibfnamefont {Nathan~C}\
  \bibnamefont {Keim}}, \ and\ \bibinfo {author} {\bibfnamefont {Sidney~R}\
  \bibnamefont {Nagel}},\ }\bibfield  {title} {\enquote {\bibinfo {title}
  {{Multiple Transient Memories in Experiments on Sheared Non-Brownian
  Suspensions}},}\ }\href@noop {} {\bibfield  {journal} {\bibinfo  {journal}
  {Phys. Rev. Lett.}\ }\textbf {\bibinfo {volume} {113}},\ \bibinfo {pages}
  {068301} (\bibinfo {year} {2014})}\BibitemShut {NoStop}%
\bibitem [{\citenamefont {Adhikari}\ and\ \citenamefont
  {Sastry}(2018)}]{Adhikari18}%
  \BibitemOpen
  \bibfield  {author} {\bibinfo {author} {\bibfnamefont {Monoj}\ \bibnamefont
  {Adhikari}}\ and\ \bibinfo {author} {\bibfnamefont {Srikanth}\ \bibnamefont
  {Sastry}},\ }\bibfield  {title} {\enquote {\bibinfo {title} {{Memory
  formation in cyclically deformed amorphous solids and sphere assemblies}},}\
  }\href@noop {} {\bibfield  {journal} {\bibinfo  {journal} {Eur. Phys. J. E}\
  }\textbf {\bibinfo {volume} {41}},\ \bibinfo {pages} {045504} (\bibinfo
  {year} {2018})}\BibitemShut {NoStop}%
\bibitem [{\citenamefont {Lilly}\ \emph {et~al.}(1993)\citenamefont {Lilly},
  \citenamefont {Finley},\ and\ \citenamefont {Hallock}}]{Lilly93}%
  \BibitemOpen
  \bibfield  {author} {\bibinfo {author} {\bibfnamefont {M}~\bibnamefont
  {Lilly}}, \bibinfo {author} {\bibfnamefont {P}~\bibnamefont {Finley}}, \ and\
  \bibinfo {author} {\bibfnamefont {R}~\bibnamefont {Hallock}},\ }\bibfield
  {title} {\enquote {\bibinfo {title} {{Memory, congruence, and avalanche
  events in hysteretic capillary condensation}},}\ }\href@noop {} {\bibfield
  {journal} {\bibinfo  {journal} {Phys. Rev. Lett.}\ }\textbf {\bibinfo
  {volume} {71}},\ \bibinfo {pages} {4186--4189} (\bibinfo {year}
  {1993})}\BibitemShut {NoStop}%
\bibitem [{\citenamefont {Deutsch}\ \emph {et~al.}(2004)\citenamefont
  {Deutsch}, \citenamefont {Dhar},\ and\ \citenamefont {Narayan}}]{Deutsch04}%
  \BibitemOpen
  \bibfield  {author} {\bibinfo {author} {\bibfnamefont {J~M}\ \bibnamefont
  {Deutsch}}, \bibinfo {author} {\bibfnamefont {Abhishek}\ \bibnamefont
  {Dhar}}, \ and\ \bibinfo {author} {\bibfnamefont {Onuttom}\ \bibnamefont
  {Narayan}},\ }\bibfield  {title} {\enquote {\bibinfo {title} {{Return to
  Return Point Memory}},}\ }\href@noop {} {\bibfield  {journal} {\bibinfo
  {journal} {Phys. Rev. Lett.}\ }\textbf {\bibinfo {volume} {92}},\ \bibinfo
  {pages} {227203} (\bibinfo {year} {2004})}\BibitemShut {NoStop}%
\bibitem [{\citenamefont {Ort{\'\i}n}(1991)}]{Ortin11}%
  \BibitemOpen
  \bibfield  {author} {\bibinfo {author} {\bibfnamefont {Jordi}\ \bibnamefont
  {Ort{\'\i}n}},\ }\bibfield  {title} {\enquote {\bibinfo {title} {{Preisach
  modeling of hysteresis for a pseudoelastic Cu-Zn-Al single crystal}},}\
  }\href@noop {} {\bibfield  {journal} {\bibinfo  {journal} {J. Appl. Phys}\
  }\textbf {\bibinfo {volume} {71}},\ \bibinfo {pages} {1454} (\bibinfo {year}
  {1991})}\BibitemShut {NoStop}%
\bibitem [{\citenamefont {Bachelard}\ \emph {et~al.}(2017)\citenamefont
  {Bachelard}, \citenamefont {Ropp}, \citenamefont {Dubois}, \citenamefont
  {Zhao}, \citenamefont {Wang},\ and\ \citenamefont {Zhang}}]{Bachelard17}%
  \BibitemOpen
  \bibfield  {author} {\bibinfo {author} {\bibfnamefont {Nicolas}\ \bibnamefont
  {Bachelard}}, \bibinfo {author} {\bibfnamefont {Chad}\ \bibnamefont {Ropp}},
  \bibinfo {author} {\bibfnamefont {Marc}\ \bibnamefont {Dubois}}, \bibinfo
  {author} {\bibfnamefont {Rongkuo}\ \bibnamefont {Zhao}}, \bibinfo {author}
  {\bibfnamefont {Yuan}\ \bibnamefont {Wang}}, \ and\ \bibinfo {author}
  {\bibfnamefont {Xiang}\ \bibnamefont {Zhang}},\ }\bibfield  {title} {\enquote
  {\bibinfo {title} {{Emergence of an enslaved phononic bandgap in a
  non-equilibrium pseudo-crystal}},}\ }\href@noop {} {\bibfield  {journal}
  {\bibinfo  {journal} {Nature Materials}\ }\textbf {\bibinfo {volume} {16}},\
  \bibinfo {pages} {808--813} (\bibinfo {year} {2017})}\BibitemShut {NoStop}%
\bibitem [{\citenamefont {Dobroka}\ \emph {et~al.}(2017)\citenamefont
  {Dobroka}, \citenamefont {Kawamura}, \citenamefont {Ienaga}, \citenamefont
  {Kaneko},\ and\ \citenamefont {Okuma}}]{Dobroka17}%
  \BibitemOpen
  \bibfield  {author} {\bibinfo {author} {\bibfnamefont {M}~\bibnamefont
  {Dobroka}}, \bibinfo {author} {\bibfnamefont {Y}~\bibnamefont {Kawamura}},
  \bibinfo {author} {\bibfnamefont {K}~\bibnamefont {Ienaga}}, \bibinfo
  {author} {\bibfnamefont {S}~\bibnamefont {Kaneko}}, \ and\ \bibinfo {author}
  {\bibfnamefont {S}~\bibnamefont {Okuma}},\ }\bibfield  {title} {\enquote
  {\bibinfo {title} {{Memory formation and evolution of the vortex
  configuration associated with random organization}},}\ }\href@noop {}
  {\bibfield  {journal} {\bibinfo  {journal} {New Journal of Physics}\ }\textbf
  {\bibinfo {volume} {19}},\ \bibinfo {pages} {053023} (\bibinfo {year}
  {2017})}\BibitemShut {NoStop}%
\bibitem [{\citenamefont {Diani}\ \emph {et~al.}(2009)\citenamefont {Diani},
  \citenamefont {Fayolle},\ and\ \citenamefont {Gilormini}}]{Diani09}%
  \BibitemOpen
  \bibfield  {author} {\bibinfo {author} {\bibfnamefont {Julie}\ \bibnamefont
  {Diani}}, \bibinfo {author} {\bibfnamefont {Bruno}\ \bibnamefont {Fayolle}},
  \ and\ \bibinfo {author} {\bibfnamefont {Pierre}\ \bibnamefont {Gilormini}},\
  }\bibfield  {title} {\enquote {\bibinfo {title} {{A review on the Mullins
  effect}},}\ }\href@noop {} {\bibfield  {journal} {\bibinfo  {journal}
  {European Polymer Journal}\ }\textbf {\bibinfo {volume} {45}},\ \bibinfo
  {pages} {601--612} (\bibinfo {year} {2009})}\BibitemShut {NoStop}%
\bibitem [{\citenamefont {Emmett}\ and\ \citenamefont
  {Cines}(1947)}]{Emmett47}%
  \BibitemOpen
  \bibfield  {author} {\bibinfo {author} {\bibfnamefont {P~H}\ \bibnamefont
  {Emmett}}\ and\ \bibinfo {author} {\bibfnamefont {Martin}\ \bibnamefont
  {Cines}},\ }\bibfield  {title} {\enquote {\bibinfo {title} {{Adsorption of
  Argon, Nitrogen, and Butane on Porous Glass.}}}\ }\href@noop {} {\bibfield
  {journal} {\bibinfo  {journal} {J. Phys. Chem.}\ }\textbf {\bibinfo {volume}
  {51}},\ \bibinfo {pages} {1248--1262} (\bibinfo {year} {1947})}\BibitemShut
  {NoStop}%
\bibitem [{\citenamefont {Middleton}(1992)}]{Middleton92}%
  \BibitemOpen
  \bibfield  {author} {\bibinfo {author} {\bibfnamefont {A~Alan}\ \bibnamefont
  {Middleton}},\ }\bibfield  {title} {\enquote {\bibinfo {title} {{Asymptotic
  uniqueness of the sliding state for charge-density waves}},}\ }\href@noop {}
  {\bibfield  {journal} {\bibinfo  {journal} {Phys. Rev. Lett.}\ }\textbf
  {\bibinfo {volume} {68}},\ \bibinfo {pages} {670} (\bibinfo {year}
  {1992})}\BibitemShut {NoStop}%
\bibitem [{\citenamefont {Preisach}(1935)}]{Preisach35}%
  \BibitemOpen
  \bibfield  {author} {\bibinfo {author} {\bibfnamefont {Ferenc}\ \bibnamefont
  {Preisach}},\ }\bibfield  {title} {\enquote {\bibinfo {title} {{\"U}ber die
  magnetische nachwirkung},}\ }\href@noop {} {\bibfield  {journal} {\bibinfo
  {journal} {Zeitschrift f{\"u}r physik}\ }\textbf {\bibinfo {volume} {94}},\
  \bibinfo {pages} {277--302} (\bibinfo {year} {1935})}\BibitemShut {NoStop}%
\bibitem [{\citenamefont {Keim}\ \emph {et~al.}(2018)\citenamefont {Keim},
  \citenamefont {Hass}, \citenamefont {Kroger},\ and\ \citenamefont
  {Wieker}}]{Keim18}%
  \BibitemOpen
  \bibfield  {author} {\bibinfo {author} {\bibfnamefont {Nathan~C}\
  \bibnamefont {Keim}}, \bibinfo {author} {\bibfnamefont {Jacob}\ \bibnamefont
  {Hass}}, \bibinfo {author} {\bibfnamefont {Brian}\ \bibnamefont {Kroger}}, \
  and\ \bibinfo {author} {\bibfnamefont {Devin}\ \bibnamefont {Wieker}},\
  }\bibfield  {title} {\enquote {\bibinfo {title} {Return-point memory in an
  amorphous solid},}\ }\href@noop {} {\bibfield  {journal} {\bibinfo  {journal}
  {arXiv:1809.08505}\ } (\bibinfo {year} {2018})}\BibitemShut {NoStop}%
\bibitem [{\citenamefont {Mukherji}\ \emph {et~al.}(2018)\citenamefont
  {Mukherji}, \citenamefont {Kandula}, \citenamefont {Sood},\ and\
  \citenamefont {Ganapathy}}]{Sood18}%
  \BibitemOpen
  \bibfield  {author} {\bibinfo {author} {\bibfnamefont {Srimayee}\
  \bibnamefont {Mukherji}}, \bibinfo {author} {\bibfnamefont {Neelima}\
  \bibnamefont {Kandula}}, \bibinfo {author} {\bibfnamefont {A~K}\ \bibnamefont
  {Sood}}, \ and\ \bibinfo {author} {\bibfnamefont {Rajesh}\ \bibnamefont
  {Ganapathy}},\ }\href@noop {} {}\bibinfo {howpublished} {arXiv:1808.07701}
  (\bibinfo {year} {2018})\BibitemShut {NoStop}%
\bibitem [{\citenamefont {Hopfield}(1982)}]{Hopfield82}%
  \BibitemOpen
  \bibfield  {author} {\bibinfo {author} {\bibfnamefont {John~J}\ \bibnamefont
  {Hopfield}},\ }\bibfield  {title} {\enquote {\bibinfo {title} {Neural
  networks and physical systems with emergent collective computational
  abilities},}\ }\href@noop {} {\bibfield  {journal} {\bibinfo  {journal}
  {Proceedings of the National Academy of Sciences}\ }\textbf {\bibinfo
  {volume} {79}},\ \bibinfo {pages} {2554--2558} (\bibinfo {year}
  {1982})}\BibitemShut {NoStop}%
\bibitem [{\citenamefont {Parisi}(1986)}]{Parisi86}%
  \BibitemOpen
  \bibfield  {author} {\bibinfo {author} {\bibfnamefont {G}~\bibnamefont
  {Parisi}},\ }\bibfield  {title} {\enquote {\bibinfo {title} {A memory which
  forgets},}\ }\href {http://stacks.iop.org/0305-4470/19/i=10/a=011} {\bibfield
   {journal} {\bibinfo  {journal} {Journal of Physics A: Mathematical and
  General}\ }\textbf {\bibinfo {volume} {19}},\ \bibinfo {pages} {L617}
  (\bibinfo {year} {1986})}\BibitemShut {NoStop}%
\bibitem [{\citenamefont {Amit}(1989)}]{amit_1989}%
  \BibitemOpen
  \bibfield  {author} {\bibinfo {author} {\bibfnamefont {Daniel~J.}\
  \bibnamefont {Amit}},\ }\href {\doibase 10.1017/CBO9780511623257} {\emph
  {\bibinfo {title} {Modeling Brain Function: The World of Attractor Neural
  Networks}}}\ (\bibinfo  {publisher} {Cambridge University Press},\ \bibinfo
  {year} {1989})\BibitemShut {NoStop}%
\bibitem [{\citenamefont {Fusi}(2017)}]{Fusi2017}%
  \BibitemOpen
  \bibfield  {author} {\bibinfo {author} {\bibfnamefont {Stefano}\ \bibnamefont
  {Fusi}},\ }\bibfield  {title} {\enquote {\bibinfo {title} {{Computational
  models of long term plasticity and memory}},}\ }\href@noop {} {\bibfield
  {journal} {\bibinfo  {journal} {arXiv.org}\ } (\bibinfo {year} {2017})},\
  \Eprint {http://arxiv.org/abs/1706.04946} {1706.04946} \BibitemShut {NoStop}%
\bibitem [{\citenamefont {Yardimci}\ and\ \citenamefont
  {Leheny}(2003)}]{Yardimci03}%
  \BibitemOpen
  \bibfield  {author} {\bibinfo {author} {\bibfnamefont {H}~\bibnamefont
  {Yardimci}}\ and\ \bibinfo {author} {\bibfnamefont {RL}~\bibnamefont
  {Leheny}},\ }\bibfield  {title} {\enquote {\bibinfo {title} {Memory in an
  aging molecular glass},}\ }\href@noop {} {\bibfield  {journal} {\bibinfo
  {journal} {EPL (Europhysics Letters)}\ }\textbf {\bibinfo {volume} {62}},\
  \bibinfo {pages} {203} (\bibinfo {year} {2003})}\BibitemShut {NoStop}%
\bibitem [{\citenamefont {Fisher}(1985)}]{Fisher85}%
  \BibitemOpen
  \bibfield  {author} {\bibinfo {author} {\bibfnamefont {Daniel~S.}\
  \bibnamefont {Fisher}},\ }\bibfield  {title} {\enquote {\bibinfo {title}
  {Sliding charge-density waves as a dynamic critical phenomenon},}\ }\href
  {\doibase 10.1103/PhysRevB.31.1396} {\bibfield  {journal} {\bibinfo
  {journal} {Phys. Rev. B}\ }\textbf {\bibinfo {volume} {31}},\ \bibinfo
  {pages} {1396--1427} (\bibinfo {year} {1985})}\BibitemShut {NoStop}%
\bibitem [{\citenamefont {Coppersmith}\ and\ \citenamefont
  {Littlewood}(1987)}]{Coppersmith87}%
  \BibitemOpen
  \bibfield  {author} {\bibinfo {author} {\bibfnamefont {S~N}\ \bibnamefont
  {Coppersmith}}\ and\ \bibinfo {author} {\bibfnamefont {P~B}\ \bibnamefont
  {Littlewood}},\ }\bibfield  {title} {\enquote {\bibinfo {title}
  {{Pulse-duration memory effect and deformable charge-density waves}},}\
  }\href@noop {} {\bibfield  {journal} {\bibinfo  {journal} {Phys. Rev. B}\
  }\textbf {\bibinfo {volume} {36}},\ \bibinfo {pages} {311} (\bibinfo {year}
  {1987})}\BibitemShut {NoStop}%
\bibitem [{\citenamefont {Cort{\'e}}\ \emph {et~al.}(2008)\citenamefont
  {Cort{\'e}}, \citenamefont {Chaikin}, \citenamefont {Gollub},\ and\
  \citenamefont {Pine}}]{Corte08}%
  \BibitemOpen
  \bibfield  {author} {\bibinfo {author} {\bibfnamefont {Laurent}\ \bibnamefont
  {Cort{\'e}}}, \bibinfo {author} {\bibfnamefont {P~M}\ \bibnamefont
  {Chaikin}}, \bibinfo {author} {\bibfnamefont {J~P}\ \bibnamefont {Gollub}}, \
  and\ \bibinfo {author} {\bibfnamefont {D~J}\ \bibnamefont {Pine}},\
  }\bibfield  {title} {\enquote {\bibinfo {title} {{Random organization in
  periodically driven systems}},}\ }\href@noop {} {\bibfield  {journal}
  {\bibinfo  {journal} {Nat. Phys.}\ }\textbf {\bibinfo {volume} {4}},\
  \bibinfo {pages} {420} (\bibinfo {year} {2008})}\BibitemShut {NoStop}%
\end{thebibliography}%

\end{document}